\documentclass[fleqn,usenatbib]{mnras}

\usepackage{newtxtext,newtxmath}

\usepackage[T1]{fontenc}

\DeclareRobustCommand{\VAN}[3]{#2}
\let\VANthebibliography\thebibliography
\def\thebibliography{\DeclareRobustCommand{\VAN}[3]{##3}\VANthebibliography}

\usepackage{graphicx}	
\usepackage{amsmath}	

\newcommand{\galfit}{{\tt{GALFIT}}}
\newcommand{\solm}{{\tt{M$_{0}$\,}}}

\title[EPOCHS Paper V. Galaxy Formation as a function of  Morphology at $z < 7$]{EPOCHS Paper V. The dependence of galaxy formation on galaxy structure at $z < 7$ from JWST observations}

\author[Conselice et al.]{
Christopher J. Conselice,$^{1}$\thanks{E-mail: conselice@manchester.ac.uk}
Justin T.F. Basham, $^{1}$
Daniel O. Bettaney,$^{1}$
Leonardo Ferreira,$^{2}$
Nathan Adams,$^{1}$
\newauthor
Thomas Harvey,$^{1}$
Katherine Ormerod,$^{1}$
Joseph Caruana,$^{3,4}$ 
Asa F. L. Bluck,$^{5}$ 
Qiong Li,$^{1}$
William J. Roper,$^{^6}$
\newauthor
James Trussler,$^{1}$
Dimitrios Irodotou,$^{7}$ 
Duncan Austin$^{1}$
\\
$^{1}$ Jodrell Bank Centre for Astrophysics, University of Manchester, Oxford Road, Manchester UK \\
$^{2}$ School of Physics and Astronomy, University of Victoria, Victoria, BC, Canada \\
$^{3}$ Department of Physics, University of Malta, Msida MSD~2080, Malta \\
$^{4}$ Institute of Space Sciences \& Astronomy, University of Malta, Msida MSD~2080, Malta\\
$^{5}$ Stocker AstroScience Center, Department of Physics, Florida International University, 11200 SW 8th Street, Miami, 33199, USA \\
$^{6}$ Astronomy Centre, University of Sussex, Falmer, Brighton BN1 9QH, UK \\
$^{7}$ Department of Physics, University of Helsinki, Gustaf Hällströmin katu 2, FI-00014, Helsinki, Finland 
}

\date{Accepted XXX. Received YYY; in original form ZZZ}

\pubyear{2024}

\begin{document}
\label{firstpage}
\pagerange{\pageref{firstpage}--\pageref{lastpage}}
\maketitle

\begin{abstract}
We measure the broad impact of galaxy structure on galaxy formation by examining the ongoing star formation and integrated star formation history as revealed through the stellar masses of galaxies at $z < 7$ based on JWST CEERS data from the Extended Groth Strip (EGS).  Using the morphological catalog of 3965 visually classified JWST galaxies from Ferreira et al. (2023), we investigate the evolution of stars, and when they form, as a function of morphological type as well as galaxies classified as passive and starburst through spectral energy distributions.  Although disk galaxies dominate the structures of galaxies at $z < 7$, we find that these disks are in general either `passive', or on the main-sequence of star formation, and do not contain a large population of starburst galaxies.   We also find no significant correlation between morphological type and the star formation rate or colours of galaxies at $z < 7$.    In fact, we find that the morphologically classified `spheroids' tend to be blue and are not found to be predominately passive systems at $z > 1.5$.  We also find that the stellar mass function for disk galaxies does not evolve significantly during this time, whereas other galaxy types, such as the peculiar population, evolve dramatically,  declining at lower redshifts.  This indicates that massive peculiars are more common at higher redshifts.  We further find that up to $z \sim 7$, the specific star formation rate (sSFR) does not vary with visual morphology, but strongly depends on stellar mass and internal galaxy mass density.  This demonstrates that at early epochs galaxy assembly is a mass-driven, rather than a morphologically-driven, process.   Quenching of star formation is therefore a mass-dominated process throughout the universe's history, likely due to the presence of supermassive black holes.  
 \end{abstract}




\section{Introduction}

For nearly a century the process of galaxy formation and evolution has remained one of the primary unsolved problems in astrophysics.  These systems, and the dark matter they contain, dominate the universe's structure, and therefore understanding them is a key part of our knowledge of how the universe came to be.  One basic question is: how (and when) do galaxies that we see in the local universe form?  This can be divided into other sub-questions, including: when did various aspects of galaxies that we see in the nearby universe come to be -- such as their structure, morphology, and masses? Similar studies have also been interested in uncovering the physical processes which have driven galaxy evolution and formation since the initial onset of star formation in the universe. In essence, the study of galaxy evolution is one of the history of galaxy properties, and then using that information to determine the physical driving mechanisms behind this observed formation \citep[][]{Bluck2014, Conselice2014b, paulino2017, osborne2020, huertas2023}.  

One aspect of this is the formation of the structures of galaxies.  This can be considered in a few ways, including the quantitative evolution of galaxies, as well as morphological `type' evolution \citep[][]{Conselice2003,Bluck2012,Mortlock2013}.   This morphological and structural evolution of galaxies can be viewed as the change in their resolved appearance and structure over time \citep[e.g.,][]{Mortlock2013,Conselice2020, Ferreira2022b, jacob2023, margalef2022}. These changes are influenced by a variety of factors, including the effects of gravity, merging, the presence of gas and dust, and the formation and subsequent evolution of stars, given a still unknown initial mass function that may vary with time \citep[][]{Trussler2022}.  All of these physical effects can drive and change the morphologies and structures of galaxies, which can then be further examined with quantitative tools \citep[e.g.,][]{Bershady2000, Conselice2003} or machine learning \citep[][]{cheng2020,cheng2021,bisigello2023}.

One of the major motivations for JWST is to understand how and when early galaxy formation occurred.  One of the main routes by which we can address this is through the structural evolution of galaxies at high redshifts $z > 3$. Specifically on account of the finite speed of light, we are viewing and imaging galaxies as they looked billions of years ago \citep[e.g.,][]{Conselice2014a, Mortlock2015, Ferreira2022}. By comparing these observations with nearby galaxies we can determine how galaxies of different types have changed throughout time.  From the earliest galaxy studies astronomers have been able to resolve these systems, and this has long provided information about the history of the formation of these objects \citep[e.g.,][]{Hubble1926}.

A major factor in the structural evolution of galaxies is the process of interactions and mergers with other galaxies. Over time, galaxies interact and merge with each other, resulting in the formation of larger, and potentially more complex structures that can have distinct morphological appearances \citep[e.g.,][]{Conselice2014a, Mortlock2013, duncan2019, Ferreira2022b, huertas2023}. This process also triggers the formation of new stars, AGN, and central massive black holes,  and the redistribution of matter within galaxies, changing appearances dramatically \citep[e.g.,][]{Conselice2003}.  Overall, the structural evolution of galaxies is a complex process that is influenced and determined by most galaxy formation processes.   As galaxies form new stars and accrete matter they continue to change and evolve over all of cosmic time.

JWST has now enabled us to investigate this problem at the highest redshifts, whereby galaxy morphology in the rest-frame optical can now be examined up to $z \sim 8$ \citep[][]{Ferreira2022, Ferreira2022b, Fudamoto2022, Kartaltepe2022}.  We can now look further into this problem, with this paper representing the next step in the process following the initial determination of morphological evolution of galaxies \citep[e.g.,][]{Ferreira2022, Ferreira2022b, Fudamoto2022, Kartaltepe2022}. One of the first major discoveries with JWST was the finding that disk galaxies dominate the morphology of galaxies all the way back to $z \sim 7$ or even to higher redshifts \citep[e.g.,][]{Ferreira2022, Treu2023}.   Before this discovery, HST observations largely pointed to peculiars as being the most common galaxy type at $z > 1.5$ \citep[e.g.,][]{Conselice2005}. In fact, it is now clear that there exists a large population of disk galaxies back to these early times, which have also been found by other studies \citep[e.g.,][]{Carnall2022, Fudamoto2022, Kartaltepe2022, Jacobs2022}.

Whilst we now have a good idea of overall galaxy morphology as a function of redshift, it is critical to investigate how galaxy formation up to $z \sim 8$ is driven by, or dependent on, galaxy structure and morphology.  We know that in the local universe there is a strong dependence of galaxy morphology on star formation and stellar mass \citep[e.g.][]{Bluck2014, Thanjavur2016}.  This correlation is such that systems classified visually as ellipticals or spheroids are the most massive galaxies which contain the lowest star formation rates \citep[e.g.,][]{Conselice2006, Kelvin2014}.  When does this relationship between structure and physical properties become established? This is now a question we can answer with JWST data, which has provided rest-frame optical structural properties of these galaxies for the first time at $z > 3$ at a higher resolution than can be provided by any previous telescope.   

Previous to JWST we thought we had a good idea of how morphology and structure evolve in terms of other galaxy physical properties, namely the star formation rate and stellar masses.   It was already clear even based on HST that does not exist a strong correlation between the morphology of a galaxy and its past and ongoing star formation rate at $z > 1.5$ \citep[e.g.,][]{conselice2011, tohill2021,papaderos2023}.   Investigations of this included looking at how stellar mass correlates with morphology up to $z \sim 3$ using HST data \citep[e.g.,][]{brinchman2000, Conselice2005, Bluck2022}.  Within this paper, we investigate this process back to the earliest times we currently can at $z < 7$, using the best morphologies that we can currently obtain with any telescope \citep[][]{Ferreira2022b}.

The paper is organized as follows. In \S~\ref{sec:data} we describe the data products we use from previous work including the visual classifications. \S~3 includes the results from the analysis examining the structures of galaxies and how they relate to star formation and stellar masses from $z=1.5$ to 7.  In \S 4 we discuss our results and what they imply for how galaxy formation occurs at $z < 8$ and  we summarise our findings from this paper in \S~5.   Throughout this paper when quanitites that require a cosmology are used (namely stellar masses and star formation rates) we use a $\Lambda$-CDM cosmology ($H_{0} = 70$ kms$^{-1}$Mpc$^{-1}$, $\Omega_{m}=0.3$ and $\Omega_{\Lambda}=0.7$).  The  values we use for these quantities are calculated using this cosmology unless explicitly stated otherwise.   This is, the measured SFR and inferred stellar masses scale as $h^{-2}$.

\section{Data and Methods}\label{sec:data}

This paper is one of series of papers from the EPOCHS survey, which is a reduction and analysis of several deep JWST imaging and spectroscopy fields.   These fields include the early Release Observations of GLASS, JADES, CEERS, and SMACS~0723, along with the PEARLS GTO Survey fields: El Gordo, MACS-0416 and the North Ecliptic Pole (NEP) \citep[][]{Windhorst2023}.  The data we use are mostly from observations taken with the \textit{Near Infrared Camera} of these various fields.  In  this paper we utilise NIRCam JWST observations using our bespoke reduction from the Cosmic Evolution Early Release Science Survey (CEERS; ID=1345).  Specifically, we use the part of this data that overlaps with the Cosmic Assembly Near-IR Deep Extragalactic Legacy Survey (CANDELS; \citealt{Grogin2011, Koekemoer2011}) in the Extended Groth Strip field (EGS).  We have reduced this data independently ourselves using a custom set-up of the \textsc{JWST} pipeline version \textsc{1.6.2} using the on-flight calibration files available through the \textsc{CDRS 0942}.  We provide an extensive description of this process and the resulting data quality in \cite{Ferreira2022b, Adams2023} and Conselice et al. (2024, in prep).  Our reduction includes all of the observed JWST bands, which includes F115W, F150W, F200W, F277W, F356W, F410W and F444W.  These bands are used in three different way thoughout this paper - for redshfit measurements, stellar masses and star formation rates, as well as for morphological measurements, both in fitting sizes as well as within visual classifications. 
    
Our analysis follows the same 3965 sources at $z > 1.5$ taken from the CANDELS catalogs, as explained in \cite{Ferreira2022b}.   This remains at the time of writing one of the largest catalogs of galaxy classifications at $z > 2$ with JWST. We use our well calibrated and robust photometric redshifts, star formation rates, and stellar masses derived for these CANDELS data  \citep{Duncan2014, duncan2019, Whitney2021} to conduct this analysis. We do not use morphological information for the selection of sources.  We also do not use magnitude cuts, as we want to ensure that sources possibly faint in HST but easily visible in the JWST observations are included. 

As explained in \citet{Ferreira2022b}, two different approaches are used to manage these data: first we perform visual classifications for all sources, which is described in detail in \citet{Ferreira2022b}. We also carry out quantitative structural measures through the code GALFIT \citep[][]{Peng2010} where parametric light profile fitting is carried out to measure sizes and Sersic indices for our sample. See \cite{Ferreira2022b, Ormerod2024} for more descriptions of how these morphologies match with the quantitative structural measurements.

\subsection{Photometric Redshifts, Stellar Masses, and Star Formation Rates}

The photometric redshifts we use in this paper originate from the redshifts calculated in \citet{duncan2019} for the Extended Groth Strip (EGS) field, although these have been checked with new JWST-based photometric redshifts using all the JWST bands mentioned above.  We use the HST based one ultimately as most of our galaxies are at lower redshifts, as well as to match the well calibrate star formation rates and stellar masses we measure.

We use the photometric redshift software \textsc{eazy} \citep{Brammer2008a} for template fitting of our SEDs, such that three separate template sets are used and fit to the photometric bands.  These templates include zero-point offsets altering the input fluxes and fixing additional wavelength-dependent errors. We also use with these redshifts a Gaussian process code (GPz; \cite{Almosallam2016}) to measure further empirical estimates using a subset of our photometric bands. Individual redshift posteriors are calculated, and all four of these measurements are combined in a statistical framework via a hierarchical Bayesian combination to create a final redshift estimate. For these quantities we use all of the HST data in this field as well as Spitzer and ground based imaging. This includes the bands F435W, F606W, F775W, F814W, F850LP, F098M, F105W, F125W, F160W, while for Spitzer the 3.6 and 4.5 $\mu$m bands, with the ground-based data including the CTIO U band, VLT/VIMOS U band, VLT/ISAAC Ks and
VLT/HAWK-I Ks data.  For a more in-depth description of the process, see section 2.4 of \citet{duncan2019}.  We use the U-band and HST based photometric redshifts as they cover most of the light where photometric redshifts are sensitive to in the redshift ranges we are probing. However, as a cross check we also examine how JWST based SEDs would differ from the ones based on HST and find essentially no difference. All the results in this paper remain the same using either version.   At $z > 3$ we find that the $\sigma_{\rm NMAD} = 0.12$ when comparing over all redshifts between the HST and JWST based photometric redshifts. 

These data used to measure these redshifts are calculated from the original CANDELS+GOODS WFC3/ACS imaging and data, Spitzer/IRAC S-CANDELS \citep{Ashby2015}, and ground-based observations with CFHT \citep{Stefanon2017} as well as with the JWST bands.  The overall method for this is described in more detail in \citet{duncan2019}.   The redshifts that we use have a high accuracy at the redshifts of interests, when compared with the spectroscopic redshifts available within the EGS field \citep[][]{duncan2019}.  
 
Our galaxy stellar masses are measured using a modified version of the spectral energy distribution (SED) code which is described in detail in \citet{Duncan2014}.   Using this code the stellar mass is measured at all redshifts in the photo-$z$ fitting range.   These masses also have what is called a `template error function', described in \citet{Brammer2008a} which accounts for uncertainties driven by the template set and any wavelength effects. 

The mass measuring technique uses \citet{Bruzual2003} templates from which we fit out data. and we include a wide range of stellar population parameters that are meant to represent the possible forms of the galaxy populations up to such high redshifts.  Our method is essentially the same as used in previous work on this sample \citep[][]{duncan2019}. To calculate these stellar masses we use a Chabrier initial mass function for the stellar populations, with a range of possible star formation histories. The star formation histories we apply follow exponential $\tau$-models for both positive and negative values of $\tau$. The timescales we use include:  $\left|\tau\right|$ = 0.25, 0.5, 1, 2.5, 5, and 10, along with a short burst model ($\tau$ = 0.05). Furthermore we also consider continuous star formation models (with $\tau \gg$ 1/$H_0$).

To ensure our masses are reliable we compare to masses calculated independently by several other teams within the CANDELS collaboration \citep[][]{Santini2015}. We do this to ensure that the stellar mass estimates do not suffer from systematic biases.  We find some minor scatter between different mass estimates at the level of 0.15 dex; however, the masses we use are not affected by any significant biases compared to others. More details on the method and models used can be found in previous work on this sample  \citet{duncan2019}.

The star formation histories we use are from the SEDs of these galaxies. We examine the UV light from these galaxies and correct for dust using the UV slope of the spectral energy distribution ($\beta$).  This gives us a measure of the dust attention in the galaxy, which we correct to obtain the total star formation rate.  These star formation rates agree well with the star formation rates derived directly from the SED fitting \citep[][]{Duncan2014}.  For more information see the papers where these measurements are described in greater detail \citep[][]{Duncan2014, duncan2019}. 

\subsection{Galaxy Classifications}

The classifications we use throughout this paper arise from the work by \citet{Ferreira2022b} who classified in total 3965 sources in the EGS field as observed with the CEERS ERS project.  These classifications were done into four main types, and were carried out by eye through the participation of five different expert classifiers.  These classifications are: disk, elliptical, peculiar, and ambiguous.  These classifications are done solely based on the F444W morphology of each galaxy based on its appearance.  No information about the stellar populations, star formation rates, or colours of the galaxies was used to determine the morphological type.   

\begin{figure*}
    \centering
    \includegraphics[width=\linewidth]{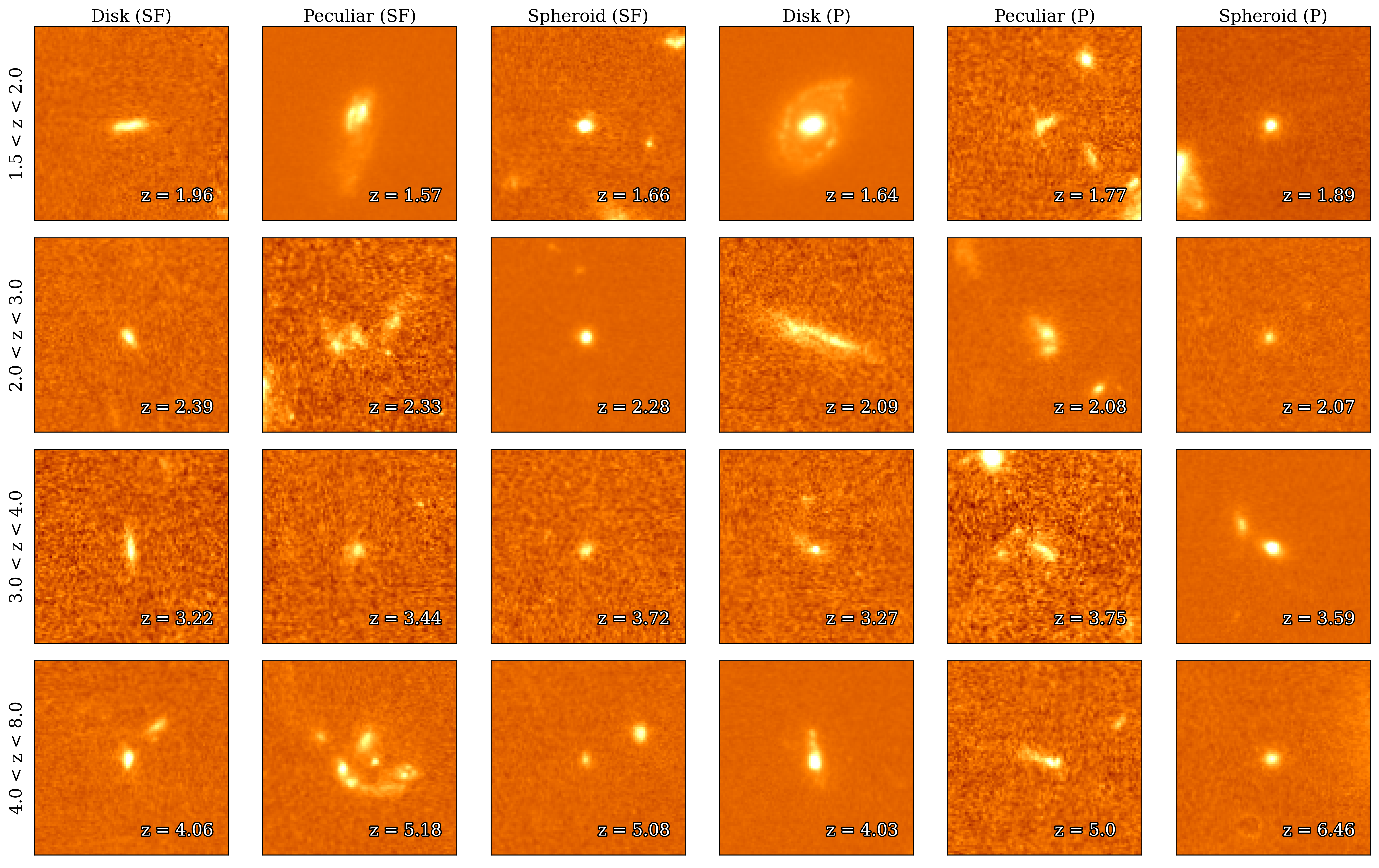}
    \caption{A subset of the images of galaxies used in this work, the image cut-outs were derived using the pipeline, outlined and described in detail in \citep[][]{Ormerod2024}. The images are ordered in increasing redshift from the top row downwards. The columns show an example of each morphology, in each redshift bin. (SF)/(P) denotes those that have been defined as star-forming/passive in the study of the main sequence. Each cut-out has a resolution of 0.03"" per pixel, and are constructed such that the background of each is scaled equally, for direct comparison. More examples are discussed and presented in \citet{Ferreira2022b}.}
    \label{fig:images}
\end{figure*}

These types are somewhat self explanatory, but we give a brief discussion here for completeness, although see \cite{Ferreira2022b} for a more detailed description of how these types are determined and the pitfalls and biases that might be present in our derived morphologies. Examples of our various types of galaxies are shown in Figure~\ref{fig:images}.  For completeness we give a brief description of how these different galaxies are classified into their various types.

First, the disk galaxies are those that appear to have a disk — this can be pure disks, edge-on disks, or systems with a central concentration (such as a bulge) with an outer disk system.  The spheroids are compact, smooth and have no obvious irregularity that changes their basic morphological structure, although very minor differences from a pure symmetry are acceptable to be considered part of this class. The 3rd main type is the peculiars, which look like galaxies in some type of active formation or disruption - essential these are analogous to those galaxies in the local universe which appear to be mergers or those dominated by star formation,  although this interpretation is not used to classify these systems.  Essentially anything that is resolved and has a morphological appearance is classified as a peculiar if it cannot be classified as an elliptical or a spiral due to the structure generally not being symmetric.  The galaxy type “ambiguous” are those galaxies that cannot be classified as they are too faint and/or too small to have any reliable classification.  Examples of these various types, besides the "ambiguous" type which are typically just small and faint featureless `blobs' are shown in Figure~\ref{fig:images} and discussed in more detail in \cite{Ferreira2022b}.

\subsection{Galaxy Size Measurements}

We examine the half light radii of our galaxies with the JWST images, using the \galfit\ software to fit a Sérsic light profile to each galaxy. \galfit\ is a least squares fitting algorithm which finds the optimum solution to a fit by using a Levenberg-Marquardt algorithm. 
Based on this, Galfit determines the goodness of fit via $\chi^{2}$ calculations, and adjusts parameters until the $\chi^{2}$ no longer improves. Galfit uses the reduced chi-squared, $\chi_{\nu}^{2}$, determined by 
\begin{equation}
\label{eqn:gal_chi}
\chi_\nu^2=\frac{1}{N_{\mathrm{DOF}}} \sum_{x=1}^{n x} \sum_{y=1}^{n y} \frac{\left(f_{\mathrm{data}}(x, y)-f_{\mathrm{model}}(x, y)\right)^2}{\sigma(x, y)^2}
\end{equation}
summed over $nx$ and $ny$ pixels, and where $N_{DOF}$ is the number of degrees of freedom. As seen in \autoref{eqn:gal_chi}, Galfit requires a data image, $f_{data}(x,y)$ and a sigma image, $\sigma(x,y)$, which are used to calculate the model image, $f_{model}(x,y)$ \citep{Galfit1, Galfit2}. 

A full description of this method can be found in \cite{Ormerod2024}, although the methodology we use to carry out these measurements is fairly straightforward.To summarise, we use a custom pipeline, taking the input parameters from the \texttt{SExtractor} catalogue, use PSFs created with \citet{Perrin2015}, and we quote the results obtained the rest-frame optical filter for each object. Many more details are presented in   \cite{Ormerod2024}. We also run our objects through another surface brightness fitting code, {\em IMFIT}, which gives very similar results.

We use the fitted sizes and Sersic indices from these measurements within this paper to investigate how quantitative structures correlate with the star formation and mass assembly of galaxies, although this is a limited study. A larger study investigating the issues of the sizes and Sersic indices of these galaxies will be discussed in \citep[][]{Ormerod2024}.

\section{Results}

In the following sections we describe the main results of this paper, especially our investigation of how galaxy properties vary as a function of morphological type in terms of measured stellar masses and star formation rates at $z < 7$.  

We first describe the stellar mass evolution and the characteristics of the evolution of galaxy stellar masses for systems at a given morphological type at these redshifts.  
We later investigate the star formation rate for these galaxies, and how the star formation rate varies with a combination of stellar mass and morphology.  

In this way we measure the dominant processes which are driving galaxy evolution at $z < 7$. 
Ultimately, we are interested in discovering what are the processes assembling galaxy mass in the universe.
This paper allows us to start investigating this question in terms of galaxy morphology at $z > 3$ for the first time with the advent of JWST data.   To do this we investigate three different quantities in this paper and how they correlate with the morphology, masses, and mass densities of galaxies.  These are ongoing star formation (SFR), the observed stellar mass (M$_{*}$), and galaxies which have since largely stopped star forming (quenched systems.).    The star formation rate can be thought of as the total mass formed within a galaxy over a certain amount of time $\delta t$, or:

\begin{equation}
    {\rm SFR_{\rm obs}} = \frac{\int_{t_0}^{t_{obs}} dM_{SFR} dt}{\delta t}
\end{equation}

\noindent where $\delta t = (t_{\rm obs} - t_{0})$, the time in which the star formation rate is being measured. As we use ultraviolet star formation rate measures, this time-scale is roughly $\delta t = 100$ Myr.  Likewise, we can the stellar mass in terms of the star formation as:

\begin{equation}
M_{*} = \int_{0}^{t_{rm obs}} {\rm SFR}(t) dt
\end{equation}

\noindent where the value of the SFR is a function of time, divided into arbitrarily small amounts.  From this equation we can also write that the SFR $\sim$ dM$_*$/dt, that is, that the SFR is the time derivative of the stellar mass.  In this sense, we use the stellar mass as the integrated formation history and the SFR as the derivative of this formation, at least averaged over the past 100 Myr.

\subsection{Galaxy measurements with redshift}

\begin{figure}
    \centering
\hspace{-0.9cm} \includegraphics[width=1.15\linewidth]{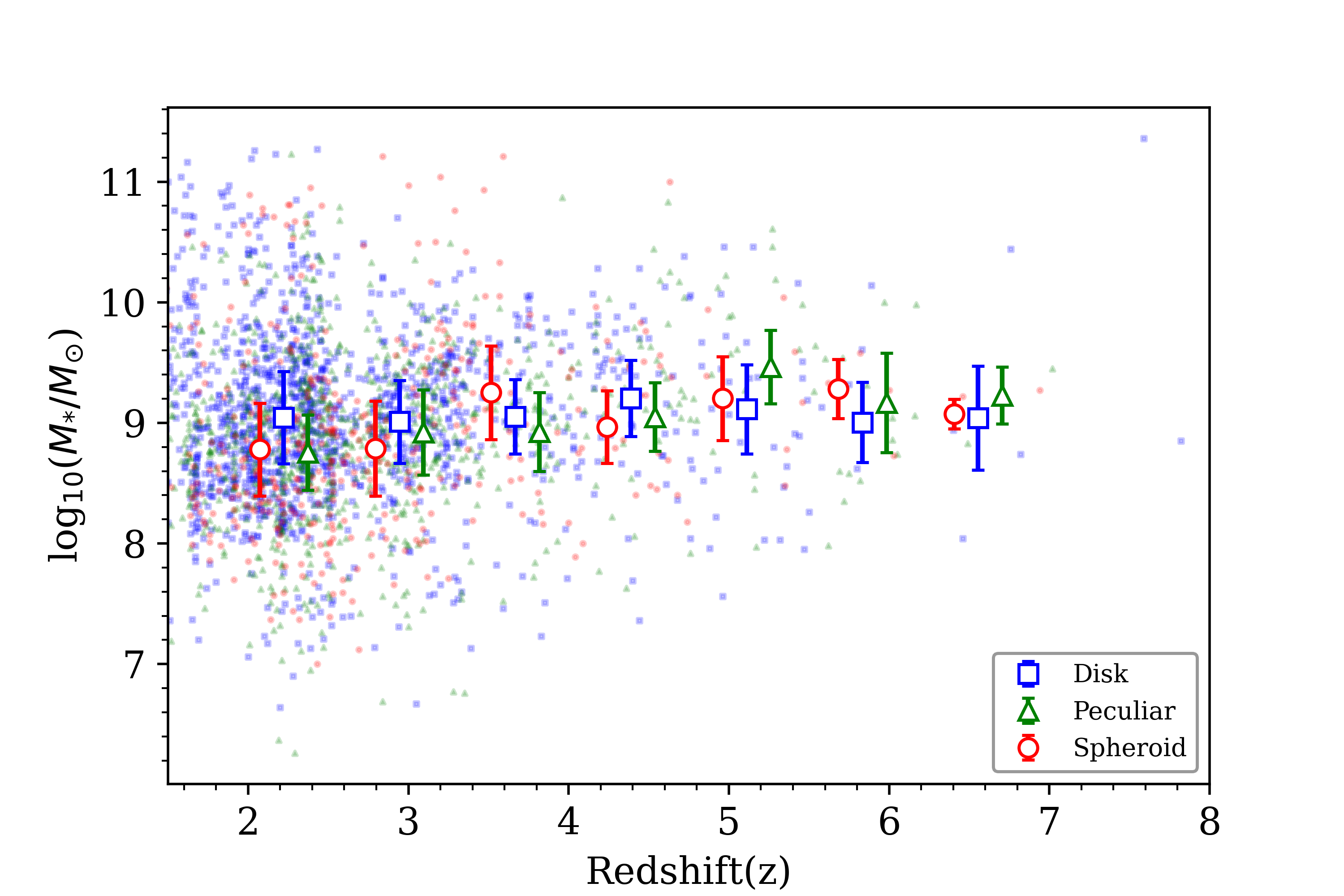}
    \caption{The plot of stellar mass as a function of redshift for our sample of galaxies with visual morphological classifications.  The different morphologies are plotted as different points, as shown in the inner caption of the figure.   In section \S 3, and throughout we investigate this relationship between the stellar mass, morphology, and star formation, and how these 
    properties vary with redshift.}
    \label{fig:massz}
\end{figure}

\begin{figure}
    \centering
\hspace{-0.9cm}    \includegraphics[width=1.15\linewidth]{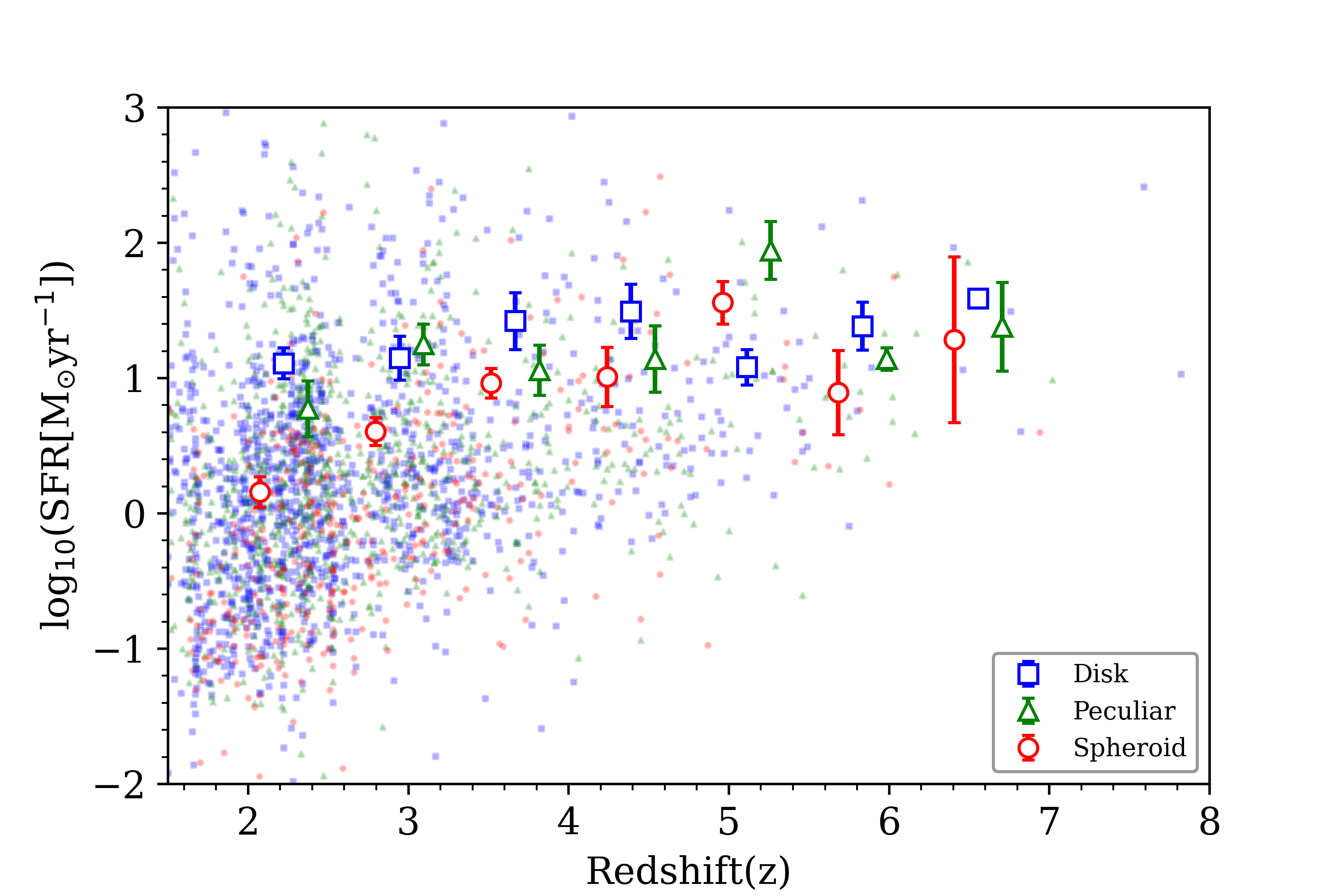}
    \caption{Plot of the scatter of the star formation rate (SFR) as a function of redshift.  The points here are coloured in terms of their morphological type, which is shown in the caption.   As can be seen, the lack of an obvious correlation between SFR and morphology can be observed.  Note that the SFR measures which are less than $\sim 1$ \solm year$^{-1}$ have errors larger than the measured value, such that there actual SFR is effectively $< 1$ \solm year$^{-1}$. }
    \label{fig:sfrz}
\end{figure}

One of the issues we investigate to get a sense of the nature of our data are the trends of star formation rate and mass with redshift for our different morphological types. 
In Figure~\ref{fig:massz} we show the scatter of distribution of stellar masses for galaxies in our sample as a function of redshift. 
As can be immediately seen, there is no obvious pattern for lower or higher mass galaxies to be of any particular morphological type.  
Likewise, we show a redshifts vs. star formation rate in Figure~\ref{fig:sfrz}, which also does not demonstrate any strong trend with any particular morphological type to be more massive or to contain higher/low star formation rates than others.    We however plot the averages of these stellar masses and star formation rates for different galaxy morphologies, showing slight differences.  One of these is that on average the spheroid types appear to have smaller star formation rates, particularly at lower redshifts. 

Based on this, and from the analysis in \citet[][]{Ferreira2022b}, we conclude that there must be multiple modes of galaxy formation occurring to produce star formation in different types of galaxies, or alternatively that different modes of star formation have different morphological outcomes as they evolve through the mechanisms responsible for triggering this star formation.
For more detail on how star formation rates correlated with galaxy types see \citet[][]{Ferreira2022b}.   This however is a very broad correlation we are finding here without any subtleties or details that can be examined from this. To understand this better we explore in more detail in the remainder of the paper how the gross morphology of a galaxy drives its past (stellar mass) and ongoing (SFR) formation history.  We carry this out by examining in some detail how galaxy morphology is distrbuted within stellar mass and star formation rates. 

\subsection{Stellar mass function evolution}
\label{sec:mass function}

One of the primary methods of examining galaxy formation is to determine the stellar mass evolution of galaxies \citep[e.g.,][]{Mortlock2011, Duncan2014, Mortlock2015,Navarro2023}.  
This quantity is the amount of stars that have formed in galaxies by a given observed time. 
Effectively it is the integral of the star formation rate at all times prior to observing the galaxy at its given redshift \citep[][]{Conselice2022, Bluck2023}.

\begin{figure*}
    \centering
    \hspace{-2cm} \includegraphics[width=1.1\linewidth]{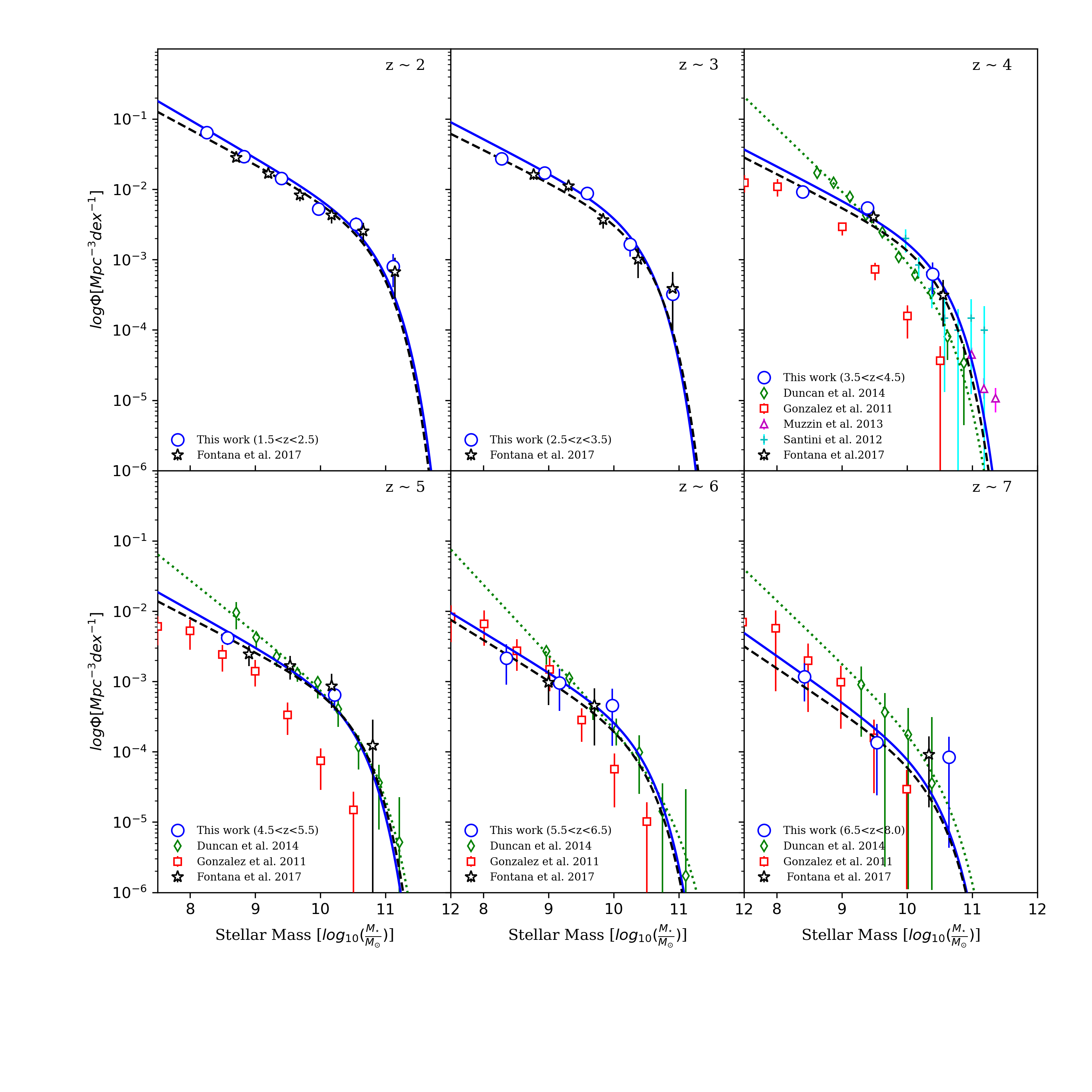}
    \vspace{-2cm}
    \caption{A comparison of our derived stellar mass functions at z $\sim $ 2 through to $z \sim 7$ with comparisons from \citet{Duncan2014}, \citet{Muzzin2013}, \citet{Gonzalez2011}, and \citet{Santini2012}. The blue (solid) line is the total mass function obtained in this work; whilst the green (dotted) line is the fit for these mass functions obtained by \citet{Duncan2014}. The black (dashed) mass function is that obtained for the galaxy sample used throughout this paper; but using the masses calculated  from similar studies \citep[e.g.,][]{Stefanon2017}}
    \label{fig:massfunall}
\end{figure*}

\begin{figure}
    \centering
    \includegraphics[width=1.1\linewidth]{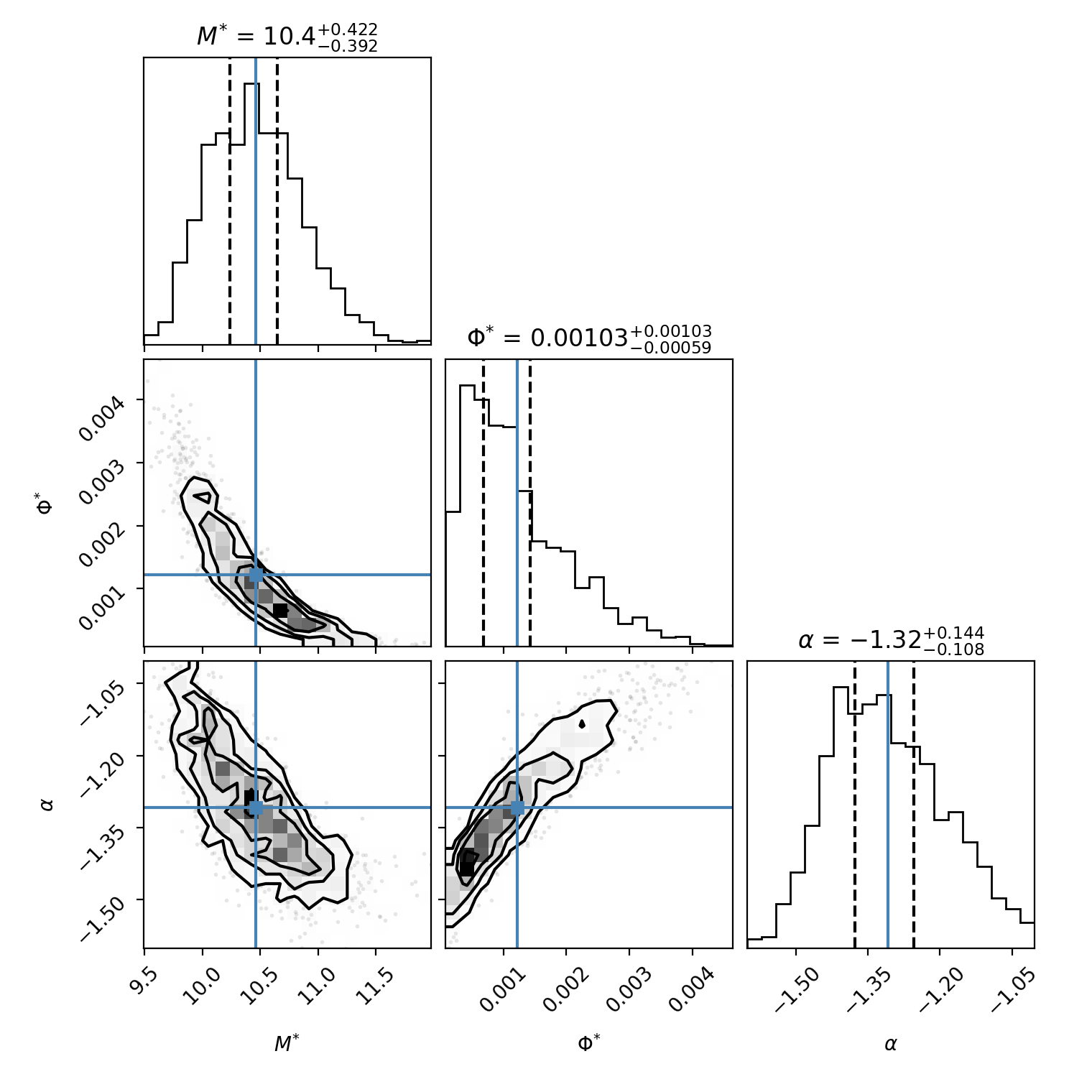}
    \caption{Corner plot from the MCMC analysis of a single mass function fit with the parameters of the fits shown in Table~2.  This is an example for the disk population in the redshift range 2.0 < z < 3.0, with similar fits carried out for the other redshift and sub-divisions into morphology ranges.}
    \label{fig:mcmc_disk_23}
\end{figure}

\begin{figure*}
    \centering
    \includegraphics[width=1.05\linewidth]{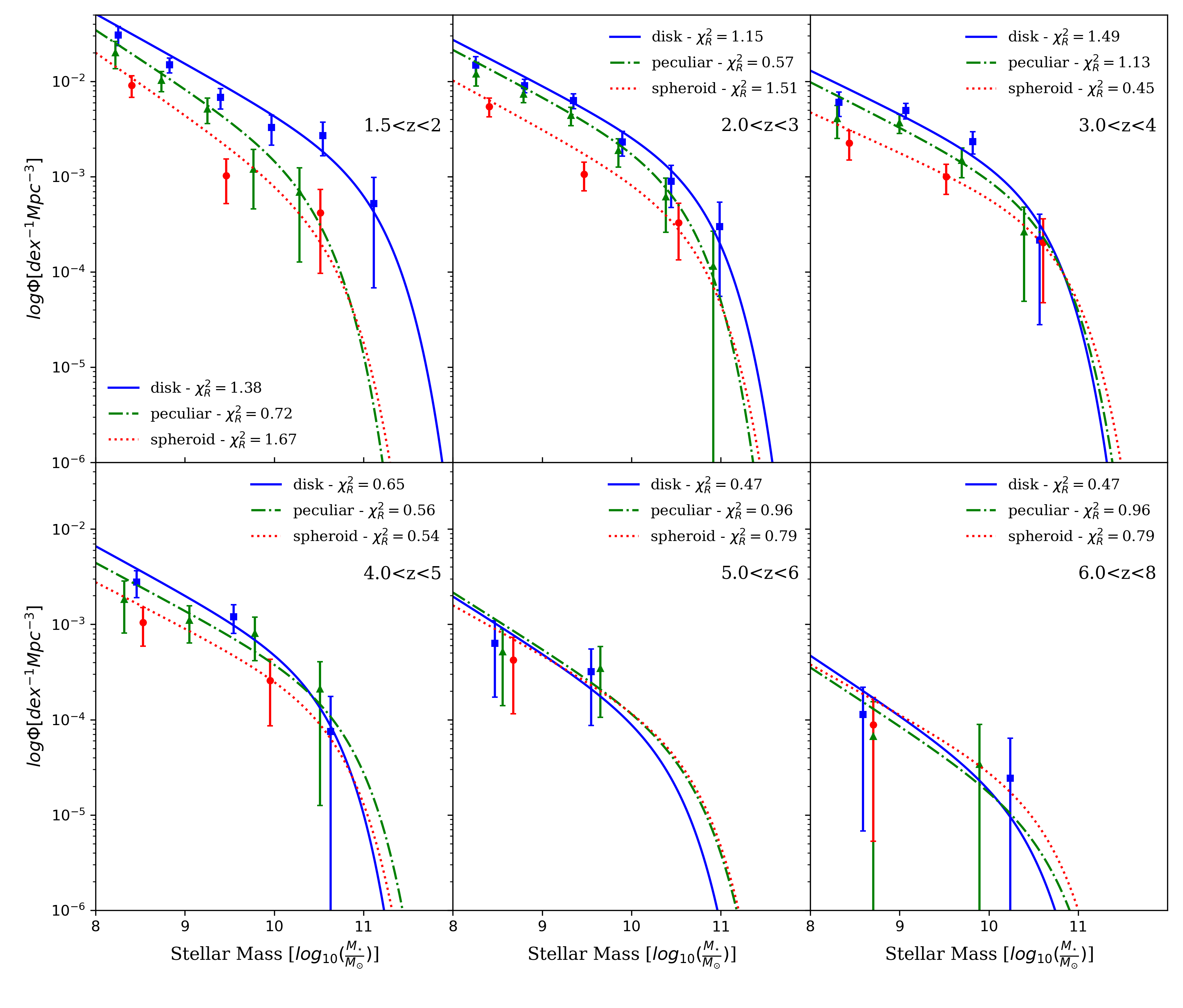}
    \caption{Derived stellar mass functions, as a function of both morphological classification type and redshift. The associated $\chi_R^2$ for the \textsc{emcee} fitting is given in the associated legend of each subplot. The lines and points are similar to what is shown in Figure~\ref{fig:massfunall}.  It is clear that at some redshifts we do not have an extensive amount of data to fit these mass functions accurately for some types, which is reflected in the limited Schechter function fits. }
    \label{fig:massfunmorph}
\end{figure*}

\begin{figure}
    \centering
    \includegraphics[width=\linewidth]{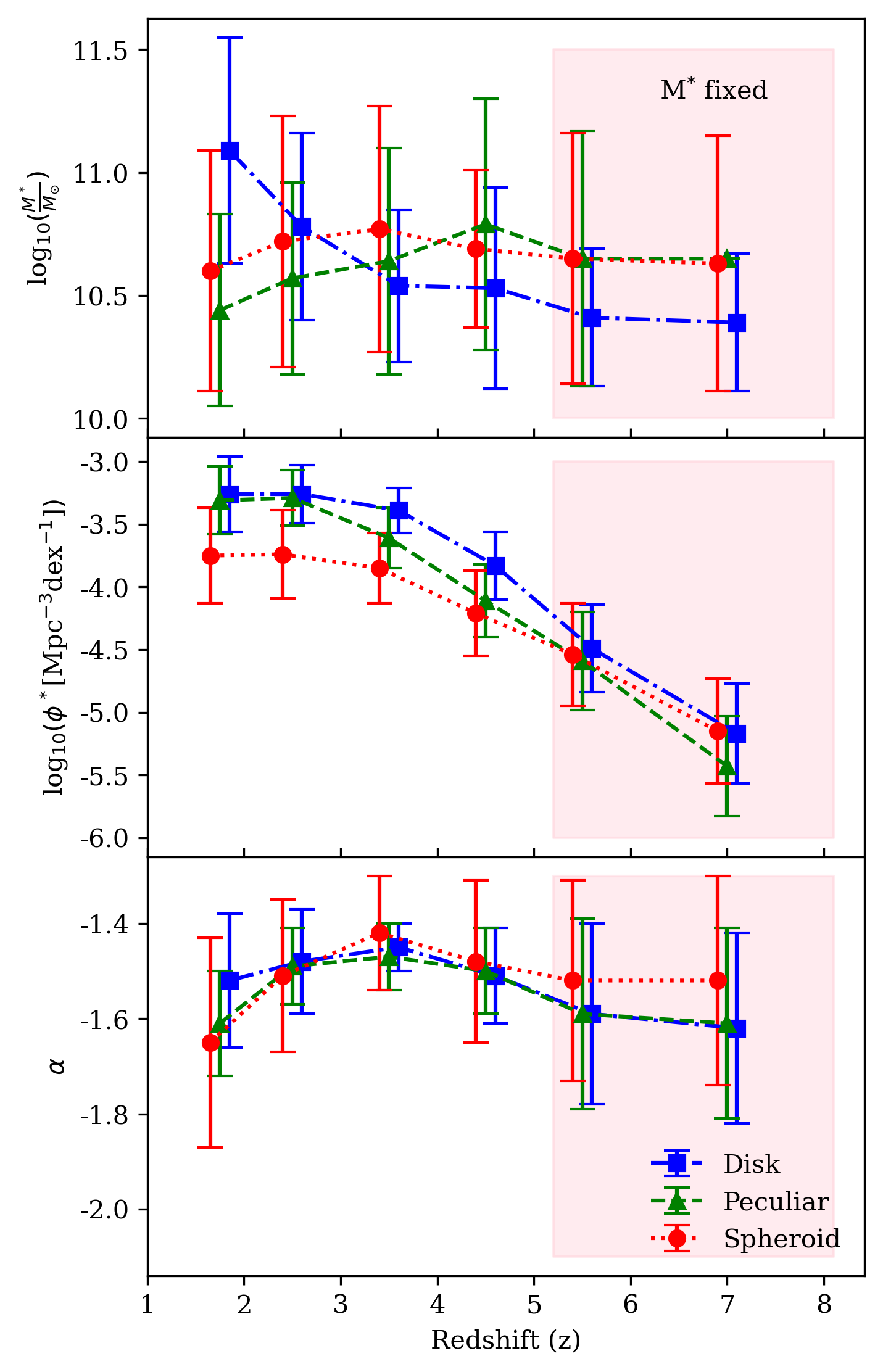}
    \caption{Visualisation of how the Schechter parameters vary through redshift as a function of morphological type. In the shaded region, at redshifts higher than the bin $ 4<z<5$, we fixed the value of $M^*$ when carrying out these fits. This is due to the turnover mass difficult to constrain given the limited amount of data.}
    \label{fig:schechter_parameters_redshift}
\end{figure}

Previous to JWST, \citet[][]{Duncan2014, duncan2019} measured the stellar masses for galaxies in the EGS field using the HST CANDELS data.  
We use these stellar masses in this paper to determine the stellar mass function evolution as a function of morphology up to $z \sim 7$, and listed in Table~\ref{tab:schechter}.  
We show our reconstruction of the total mass function in Figure~\ref{fig:massfunall}.  
We recalculated the \cite{Schechter1976} function fit parameters using our fitting methodology, which is described in detail below, and we find a good agreement with the previous work on this topic from \citet[][]{Duncan2014} amongst others, as can be seen in Figure~\ref{fig:massfunall}.

\begin{table}
\begin{tabular}{ | c | c | c | c  | c | }
\hline
Redshift Range & $M^{*}$ & $\log \phi_{*}$ & \(\alpha\) & $\chi_{R}^2$ \\ 
\hline
$1.5<z<2$  & 10.86 $\pm$ 0.17 & -2.92 $\pm$ 0.18 & -1.54 $\pm$ 0.18 & 1.16 \\
$2<z<3$  & 10.40 $\pm$ 0.21 &-2.80 $\pm$ 0.17 & -1.48 $\pm$ 0.16 & 1.63 \\
$3<z<4$  & 10.51 $\pm$ 0.24 & -3.24 $\pm$ 0.18 & -1.48 $\pm$ 0.16 & 2.40\\
$4<z<5$  & 10.51 $\pm$ 0.24 &  -3.65 $\pm$ 0.22 & -1.52 $\pm$ 0.14 & 0.88\\
$5<z<6$  & 10.42 $\pm$ 0.32 & -4.02 $\pm$ 0.30 & -1.56 $\pm$ 0.16 & 0.32\\
$6<z<7$ &  10.39 $\pm$ 0.28 & -4.55 $\pm$ 0.34 & -1.65 $\pm$ 0.20 & 0.77 \\
\hline
\end{tabular}
\centering
\caption{The single Schechter function fitted parameters for the total galaxy stellar mass function for our whole data set, encompassing all morphology types.  Shown are the error bars for these fits which were derived using our MCMC fitting methods. Our results agree well with previous mass function measurements, including from \citet[][]{Duncan2014}.}
\label{tab:schechter_all}
\end{table}

\begin{table*}
\begin{tabular}{ | c | c | c | c | c  | c | }
\hline
Redshift Range &  Type & $M^{*}$ & $\log \phi_{*}$ & \(\alpha\) & $\chi_{R}^2$ \\ 
\hline
$1.5<z<2$ & Disk & 11.09 $\pm$ 0.46 & -3.26 $\pm$ 0.30 & -1.52 $\pm$ 0.14 & 1.38 \\
$2<z<3$ & Disk & 10.78 $\pm$ 0.38 &-3.26 $\pm$ 0.23 & -1.48 $\pm$ 0.11 & 0.98 \\
$3<z<4$ & Disk & 10.54 $\pm$ 0.38 & -3.39 $\pm$ 0.23 & -1.45 $\pm$ 0.12 & 1.20\\
$4<z<5$ & Disk & 10.53 $\pm$ 0.41 &  -3.83 $\pm$ 0.27 & -1.51 $\pm$ 0.10 & 0.65\\
$5<z<6$ & Disk & 10.41 $\pm$ 0.28 & -4.49 $\pm$ 0.35 & -1.59 $\pm$ 0.19 & 0.47\\
$6<z<8$ & Disk & 10.49 $\pm$ 0.00 (Fixed) & -4.75 $\pm$ 0.31 & -1.58 $\pm$ 0.17 & 0.21\\
$6<z<8$ & Disk & 10.39 $\pm$ 0.28 & -5.17 $\pm$ 0.28 & -1.62 $\pm$ 0.20 & 0.40\\
$1.5<z<2$ & Peculiar & 10.44 $\pm$ 0.39 & -3.31 $\pm$ 0.27 & -1.49 $\pm$ 0.11 & 0.72 \\
$2<z<3$ & Peculiar & 10.57 $\pm$ 0.39 &-3.29 $\pm$ 0.22 & -1.49 $\pm$ 0.08 & 0.57 \\
$3<z<4$ & Peculiar & 10.64 $\pm$ 0.46 & -3.61 $\pm$ 0.24 & -1.47 $\pm$ 0.07 & 1.13\\
$4<z<5$ & Peculiar & 10.79 $\pm$ 0.51 &  -4.11 $\pm$ 0.29 & -1.50 $\pm$ 0.09 & 0.56\\
$5<z<6$ & Peculiar & 10.65 $\pm$ 0.52 & -4.59 $\pm$ 0.39 & -1.59 $\pm$ 0.20 & 0.96\\
$6<z<8$ & Peculiar & 10.65 $\pm$ 0.00 (Fixed) & -5.43 $\pm$ 0.40 & -1.61 $\pm$ 0.20 & 0.29\\
$1.5<z<2$ & Spheroid & 10.60 $\pm$ 0.49 & -3.75 $\pm$ 0.38 & -1.65 $\pm$ 0.22 & 1.67 \\
$2<z<3$ & Spheroid & 10.72 $\pm$ 0.51 & -3.74 $\pm$ 0.35 & -1.51 $\pm$ 0.16 & 1.51 \\
$3<z<4$ & Spheroid & 10.77 $\pm$ 0.50 & -3.85 $\pm$ 0.28 & -1.42 $\pm$ 0.12 & 0.45\\
$4<z<5$ & Spheroid & 10.69 $\pm$ 0.32 &  -4.21 $\pm$ 0.34 & -1.48 $\pm$ 0.17 & 0.54\\
$5<z<6$ & Spheroid & 10.65 $\pm$ 0.51 & -4.54 $\pm$ 0.41 & -1.52 $\pm$ 0.21 & 0.79\\
$6<z<8$ & Spheroid & 10.63 $\pm$ 0.52  & -5.15 $\pm$ 0.42 & -1.52 $\pm$ 0.22 & 0.84\\

\hline
\end{tabular}
\centering
\caption{The single Schechter parameters for the total galaxy stellar mass function as a function of morphological type. Shown are the fitted parameters assuming that the mass function is well fit by this function.   We also note on this Table when the value of $M^{*}$ is held constant when the other Schechter function parameters are fit.   We also list the value of the reduced-chi-squared for each fit ($\chi^{2}_{R}$).}
\label{tab:schechter}
\end{table*}

We fit the Schechter function parameters for our sample of galaxies using the following procedure. 
We compute the number densities of galaxies as the number of galaxies in a given stellar mass bin normalised by the co-moving volume in the corresponding redshift bin. 
We also correct our counts for incompleteness using the results from \citet{duncan2019}, analysing the completeness of the EGS Wide 1 subfield.
These inferred mass functions are then fit by creating an array of all the computed number densities in order of the centre of the stellar mass bin. 
We then fit for each bin a Schechter function (\citealt{Schechter1976}) of the form:

\begin{equation}
\phi(M) = \phi^{*}\cdot \rm \ln(10)\cdot[10^{(M-M^{*})}]^{(1+\alpha)}\cdot \rm \exp[-10^{(M-M^{*})}]
\label{eq:sch}
\end{equation}

\noindent the value $\phi^{*}$ is the normalisation, $M^{*}$ is the value of the so-called turn-over mass in dex units, while $\alpha$ is the slope of the low-mass end distribution.\\

\noindent The parameter determination was conducted via use of the  Markov Chain Monte Carlo (MCMC) sampler \textsc{emcee} \citep{Foreman-Mackey2016}, in which we take the likelihood function to be of Gaussian form, and maximise this with a uniform prior. 
The sampling of the posterior is done using \textsc{emcee} to draw probable parameter arrays from the posterior probability distribution. 
We experiment with other prior distributions, finding the outcome of these results very similar to the main process we describe here.
This method is visualised using the plotting software \textsc{corner} \citep{corner}. 
For example, histograms of the posterior distribution for the disk galaxy population between $1.5< z < 2$ are shown in Figure.~\ref{fig:mcmc_disk_23}.   
The other morphological types show similar features and similar corner plots from our MCMC fitting of these mass functions.   The best fits for these mass functions is shown in Table~\ref{tab:schechter} as a function of morphology, while the total mass function fits are shown in Table~\ref{tab:schechter_all}. 

At the highest redshifts we are limited to a sample of a few dozen galaxies for each of the three major morphological types, as well within the lower redshift bins, there are a few hundred. These limited numbers should be kept in mind when interpreting the results and trends that we find in this paper.  

\subsection{Mass functions of morphological types}

We now examine how the different morphological types of our sample are distributed as a function of stellar mass.  
This is shown in Figure~\ref{fig:massfunmorph}, where as a function of redshift we have plotted the stellar mass function for galaxies in the classified morphological types of disks, peculiars, and spheroidal galaxies.   
The Schechter function parameters for each of the morphological types are listed as a function of redshift in Table~\ref{tab:schechter}.   
The fitted Schechter function parameters are also plotted in Figure~\ref{fig:schechter_parameters_redshift}.
Note that in several versions of the fits, we hold the values of M$^{*}$ constant, as otherwise we do not have enough data to fully fit these. As can also be seen, it becomes progressively harder to measure accurately points at the highest redshifts (Figure~\ref{fig:massfunmorph}).  These mass functions will be better measured when larger morphological samples of galaxies become available.  

The largest difference we see within these fits, is the high density of disk galaxies as shown by the values of $\phi$ amongst all the redshifts.  
As shown in Figure~\ref{fig:schechter_parameters_redshift}, the disk galaxies have a higher normalisation than any other morphological types at all redshifts, which is another reflection of the fact that the disks dominate the galaxy population at these redshifts \citep[][]{Ferreira2022b}.
The spheroid galaxies have the lowest densities, particularly at the lower redshifts. 
However at $z >  5$ we find that the values of $\phi$ are fairly similar for all morphological types. 
One caveat to this is that we have held the value of M$^{*}$ constant at these highest redshifts which may affect the fits of the other parameters, although the value of M$^{*}$ is fairly constant at the higher redshifts \citep[e.g.,][]{Duncan2014, Harvey2024} so the effect is likely small. 
Interestingly, we find that the values of $\phi$ grow at very similar rates for the different morphological types. It is not clear however if between these redshifts we are seeing a growth in the number of galaxies that, for whatever reason, are distributed equally between types, but we cannot determine from this work alone how the morphologies between redshifts might be evolving into each other.  

Remarkably perhaps, we find no significant difference in the value of the faint-end slope, $\alpha$ for the galaxies classified into different morphological types.
This however is more or less only the case at the highest redshifts at $3 < z < 5$.
At the lowest redshifts we find that there are fewer lower-mass spheroid galaxies, and a relatively higher number of disks and peculiars at the faint end of the mass function.

\subsection{Evolution of the Main Sequence (MS): Passive, Starbursts and MS galaxies}

\subsubsection{Definitions}

Galaxies come in different types, not only morphological, but also in terms of different ongoing star formation rates.  
Some galaxies are undergoing star formation in a burst mode, some are passive (with little star formation), whilst others are present in what is called the `main-sequence' of star formation.    It is clear that especially at high redshifts, there is no obvious correlation between the morphology of a galaxy and its star formation rate \citep[e.g.,][]{conselice2011, Mortlock2013}. However, a strong dependence of sSFR on quantitative structure has been found, including bulge mass, central mass density and the stellar potential \citep[e.g.][]{Cheung2012, Fang2013, Lang2014, Bluck2022, Bluck2023}. It is interesting that these structural relations do not appear to result in strong morphological dependence on star formation.

As part of this we investigate the so-called "main-sequence" of galaxy formation for our sample and how morphological properties relate to this. 
This main-sequence is a (sub-)linear relationship between the star formation rate of galaxies and their past integrated star formation as demonstrated by their measured stellar mass  \citep[e.g.,][]{Noeske2007, Bhatawdekar2021}.  These systems can be thought of as `normal' galaxies that are neither passive nor star bursting.  We use our data to define passive, starbursting, and main-sequence types at each redshift galaxies by fitting the main sequence and determining the different star forming types based on the position of individual galaxies relative to this fit, with star forming galaxies defined as those systems which are $> 1 \sigma$ above the main sequence, and passives those which are $< 1 \sigma$ below the main sequence.

Our definition of the main sequence originates via the assumption of a simple power law relation between star formation rate and the stellar mass \citep[][]{mainsequence2018}. 
A straight line fit to the $\log_{10}($SFR$)$ vs $\log_{10}$(M$_*$) data is performed using \textsc{emcee}, wherein the vertical scatter of each point from this straight line is found to be normally distributed. 
Galaxies within one vertical standard deviation of this fit were then isolated for fitting via \textsc{emcee}. This produces a final straight line that defines our main sequence, of the form:

\begin{equation}
\log_{10}\left(\frac{\textrm{SFR}}{\textrm{M}_{\odot}\textrm{yr}^{-1}}\right) = \gamma\log_{10}\left(\frac{\textrm{M}_*}{\textrm{M}_{\odot}}\right) + \beta .
\label{eq:mainsequence}
\end{equation}

\noindent With respect to this line, each galaxy's $\log_{10}($SFR$)$ has a vertical deviation ($\delta$) that is normally distributed across the population. 
The width of this distribution ($\sigma$) was calculated, such that the starbursting, main sequence, and passive populations are defined as:\\

\centerline{Starbursting : $ \delta > \sigma $} 
\centerline{Main Sequence : $ \vert \delta \vert <  \sigma $} 
\centerline{Passive : $ \delta < -\sigma $} 

\noindent This procedure was repeated within redshift bins at z = (1.5, 2, 3, 4, 7), providing the fractional starforming history makeup of the starbursting, main sequence, and passive populations as a function of redshift.To be clear, we carry out this fit to determine this criteria for all the galaxies within a redshift bin; we not fit this for the individual morphological types.  Thus the fits shown in the overall stellar mass vs. SFR plane are the fits over all galaxies within that redshift bin.

\begin{figure*}
    \centering
    \includegraphics[width=1.0\linewidth]{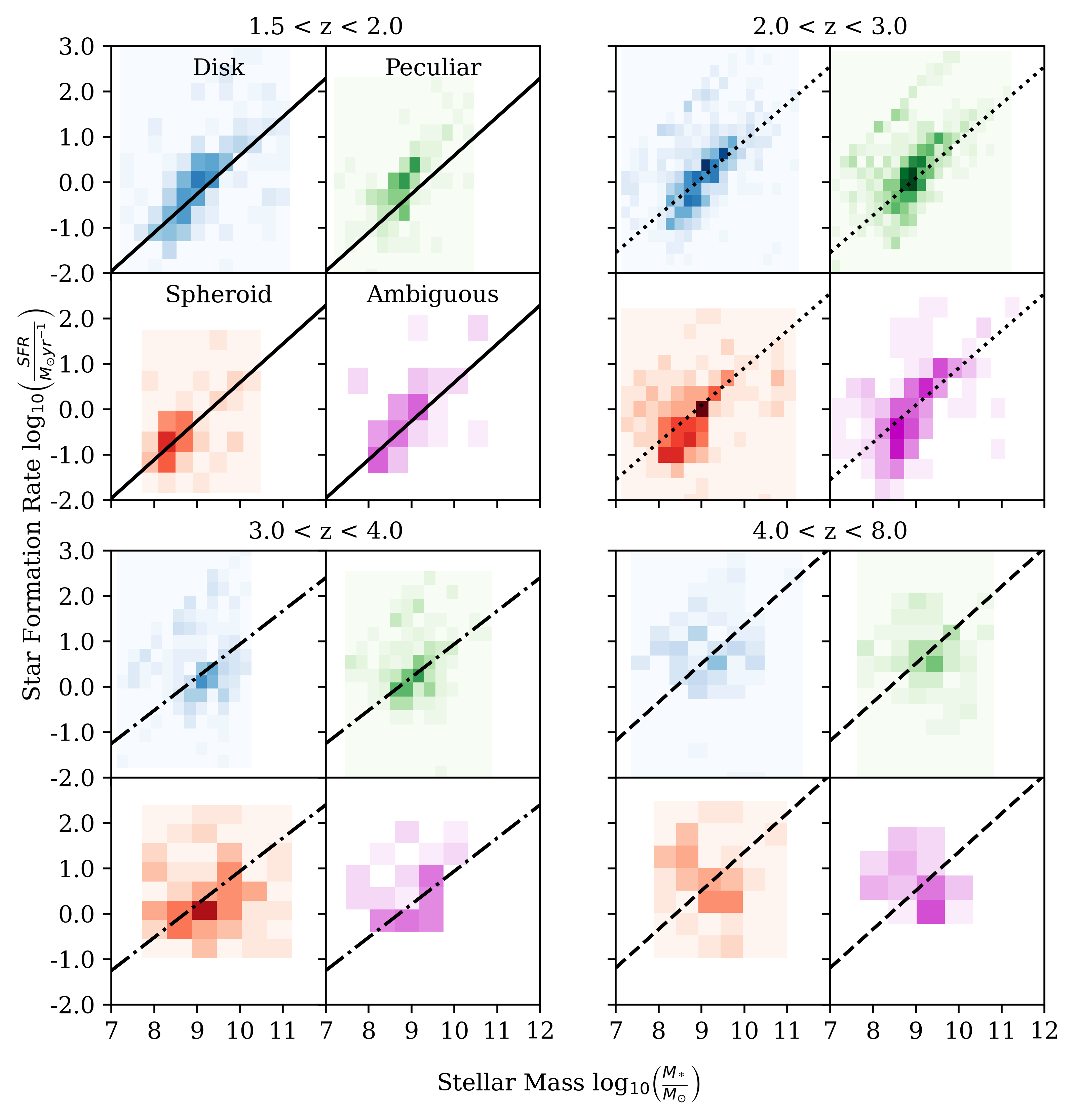}
    \caption{Plot showing the densities of galaxies in the main sequence plane by morphological type, wherein the over-plotted straight in each redshift bin has been obtained via a fit to ourSFR and M$_{*}$ values.  The colours of the symbols show the morphological type of the galaxies plotted.  The blue points are for the disk galaxies, the red are those for the spheroids, the green are peculiars, and the ambiguous galaxies are plotted as purple.  }
    \label{fig:main_sequence}
\end{figure*}

\subsubsection{Main-sequence types}

The results of this exercise are shown in Figure~\ref{fig:main_sequence} where we have divided up the main sequence fitting into different redshift and morphological bins.  There are a few things to take away from this figure.  The first is that there is well defined main sequence for galaxies at most morphologies, with the exception of the classified spheroids, up to $z \sim 3$.  This can be seen as the line showing the best fit to the data.   It is also the case that there is not a great deal of difference in the main sequence fitting and distribution of galaxies in terms of morphology up to $z \sim 3$.  In the upper panels of Figure~\ref{fig:main_sequence} it can be seen clearly that the distribution is very similar and that there are no ‘offsets’ for the main sequence at $1.5 < z < 3$. This also means that the main sequence is made up of all morphology types in this epoch, which is a bit different from the local universe where it is found in general that the morphological make up of the main sequence is primarily from disk galaxies \citep[e.g.,][]{Brinchmann2004,Noeske2007}.

At the higher redshifts at $z > 3$, where we now have the first reliable morphological types, we do not see as clear of a pattern.  The main sequence is still present to some degree, but there is a larger scatter of points.  None the less, these points seem to fall into the same part of the parameter space for the different galaxy morphologies.  

\begin{figure*}
    \centering
    \includegraphics[width=\linewidth]{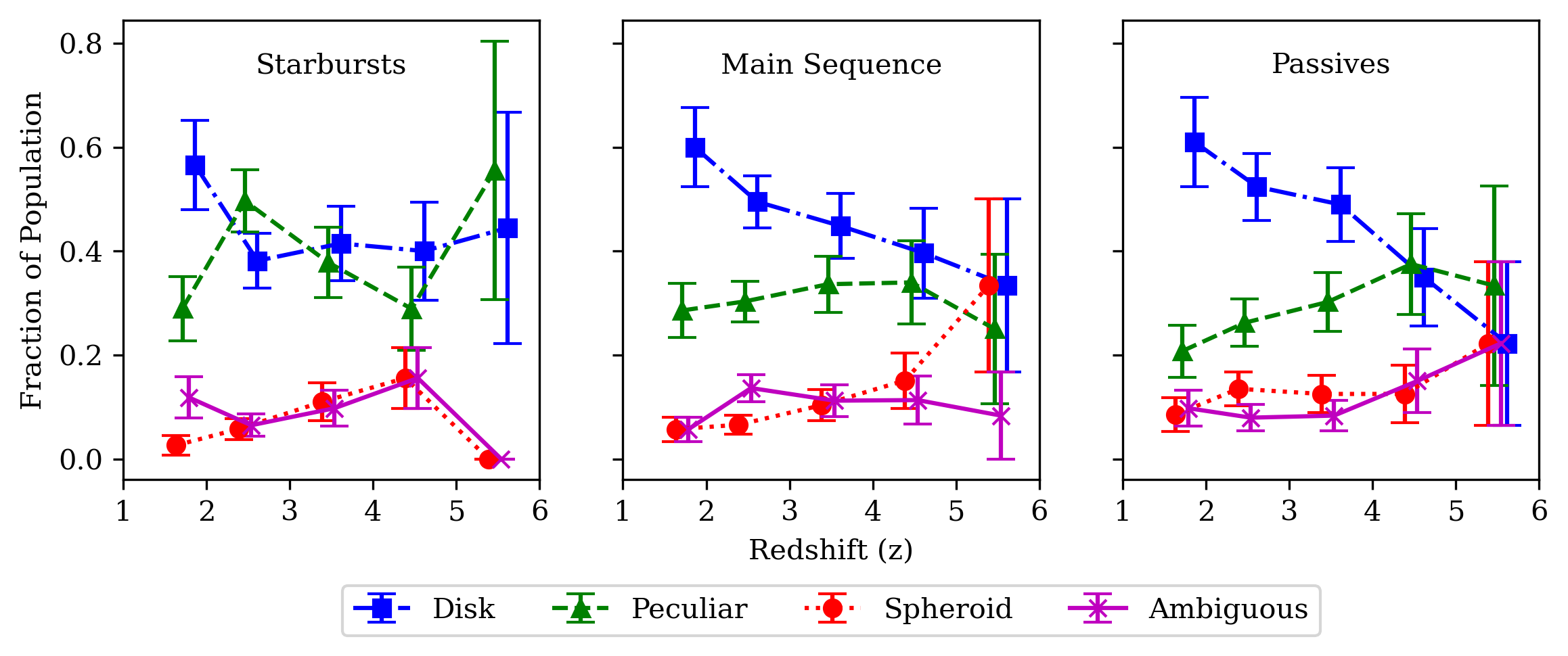}
    \caption{The morphological fractional composition of the ''main-sequence" method of defining star forming types, which are: starbursting, main sequence, and passive galaxies divided by visual morphology as a function of redshift.  Plotted are the fraction of galaxies defined as starbursts, main sequence, and passive as a function of redshift.  The different colours and symbols show the breakdown of the fraction of each star formation type as a function of redshift.  Compare this to Figure~\ref{fig:fractions_morph} where these morphological main-sequence types are plotted in terms of morphological type.}
    \label{fig:fractions_ms}
    \end{figure*}

\begin{figure*}
    \centering
    \includegraphics[width=0.8\linewidth]{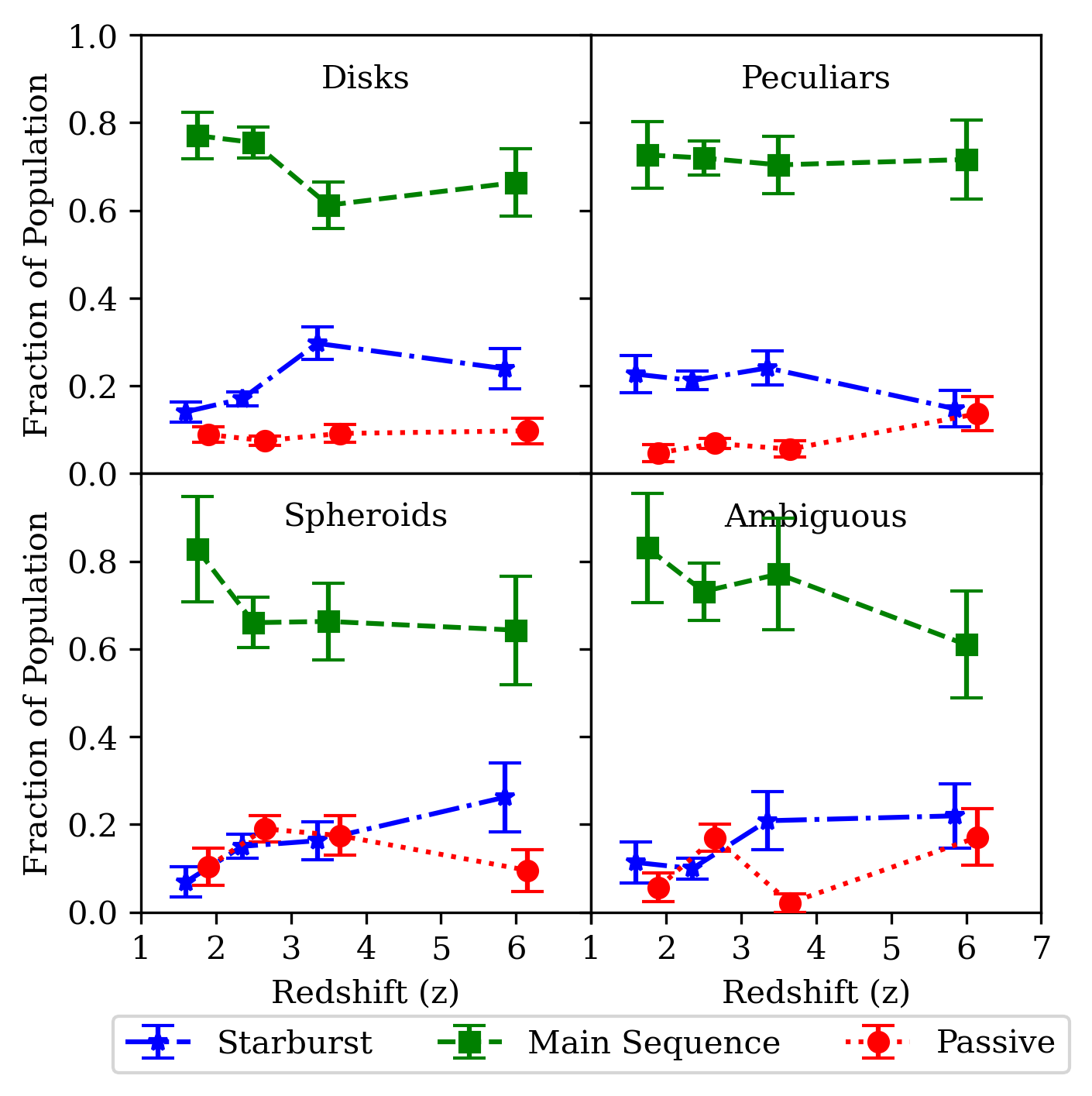}
    \caption{The main-sequence defined star formation types fractional composition of the disk, peculiar, spheroidal and ambiguous morphologically classified galaxies by star forming category as a function of redshift. The different colours and symbols show the breakdown of the fraction of each morphology within a given star forming type as a function of redshift.}
    \label{fig:fractions_morph}
    \end{figure*}

The best way to determine how the different galaxy types change along the main sequence, or are defined as passive or starburst is to plot the change in the morphological types as a fraction of these main sequence based derived star formation states. The results of this are shown in Figure~\ref{fig:fractions_ms} and Figure~\ref{fig:fractions_morph}.  It is important to note how many galaxies we have within each of these morphological bins at different redshifts.  At $z = 1.5-2$ we have 507 disk galaxies, 268 peculiar galaxies, and 149 spheroid galaxies within these plots.  This changes at $z \sim 3$ to be 463 peculiars, 619 disks, and 214 spheroids, while at $z \sim 6.5$ we are left with 29 peculiars, 34 disks, and 16 spheroids.

To determine which star formation type a galaxy is we use its distance from the main-sequence, which we determine by fitting within each redshift bin, all galaxies, regardless of morphology, to defined the main-sequence.  Then the various criteria are applied in reference to this fit. 
The first plot in Figure~\ref{fig:fractions_ms} shows the fraction of the star forming types as a function of morphological type — the starbursts, main-sequence galaxies and those which are passives.  We divide up this trend in this way as we know from our definition that a large fraction of all galaxies will naturally be defined as main-sequence using our definition. However, this does not mean that there should be any particular biases for any morphological type of galaxy to be found on the main-sequence or within the passives or starbursts.

As deduced from the plot of the main sequence itself, we find a mixture of galaxy morphologies that make up the main-sequence, although we do see some evolution.  At the highest redshifts where we can carry out this test, at $z \sim 6$, we find that the relative contributions of spheroid, disk and peculiar are fairly similar, at about a third of the population each, with spheroids dropping quickly at the next lower redshift bin to be only about 10\% of the population.  At lower redshifts we increasingly see that the disk galaxies make up a large fraction of the main-sequence galaxy population, such that it constitutes about 60\% of the main sequence at $z \sim 1.5$.      This compares well with the total fraction of disk galaxies that make up the entire galaxy population at this redshift \citep[][]{Ferreira2022b}, demonstrating again that disk galaxies are dominating the typical star formation in the universe at all epochs up to $z \sim 6$.    

Peculiar galaxies have a slightly declining contribution to the fraction of galaxies on the main-sequence, whilst the spheroids continue to have a lower contribution and become, on average, more passive.  This is essentially evidence that we are witnessing the formation of the Hubble sequence.  Galaxies must by nature form in some type of irregular/peculiar way and transform into normal galaxies at lower redshifts. Thus, we are finding at the lowest redshifts the expected and typical trend of galaxy morphology correlating with the star forming properties of galaxies with different morphologies.

The starburst population, those defined as above the main sequence, are equally dominated by disk galaxies and peculiars.  In fact these are about equally represented, while ellipticals again represent a very small fraction of this population.  On the other hand, the ‘passive’ types are seen to have more similarities to the ‘main-sequence’, such that the three types are fairly equal at the highest redshifts.  However at lower redshifts we see that the disk galaxies become the dominant type, again at about 60\% of the population at the lowest redshifts.  The major difference between this and the main sequence galaxies is that the peculiars decline more rapidly — there are in fact very few passive peculiar galaxies at the lowest redshifts.  

We can reverse this view and ask the question in terms of the morphological types, essentially - what fraction of each type is classified within the different star formation categories as based on the main sequence?  This is shown in Figure~\ref{fig:fractions_morph} where the dissection of each morphological type is shown in terms of these star forming properties.   The trends here reflect what we see when we examine the evolution of the star forming types in terms of morphology.

The disk galaxies display a high fraction of their number on the main-sequence, which by their dominance and the definition of our main sequence, is not a surprise.  However, we can see that there is gradual evolution in terms of the fraction, such that they evolve away from being starbursts and slightly more on the main-sequence, particularly at redshifts $z < 3$.  However, it is still the case that even at redshift $z \sim 6$, 60\% of disk galaxies are on the main sequence.  Spheroids show similar trends, and again this is consistent with a gradual formation of the Hubble sequence, although the early onset of this is actually at redshifts higher than our highest redshift at $z > 6$.  Deeper and more imaging data are needed to probe this in more detail at these earlier times, which might be possible with MIRI and deep NIRCam observations.  
Peculiars show different trends, whereby they increase slightly their fraction in the starburst class at lower-z, while retaining a fairly flat evolution with redshift in the main-sequence.

\begin{table}
\begin{tabular}{ | c | c | c | c | }
\hline
Redshift Range &  $\gamma$ & $\beta$ & $\sigma$\\ 
\hline
$1.5 < z < 2.0$ & 0.85 $\pm$ 0.22 & -7.91 $\pm$ 2.09 & 0.62 \\
$2.0 < z < 3.0$ & 0.82 $\pm$ 0.22 & -7.29 $\pm$ 1.98 & 0.69 \\
$3.0 < z < 4.0$ & 0.73 $\pm$ 0.18 & -6.36 $\pm$ 1.68 & 0.72 \\
$4.0 < z < 8.0$ & 0.85 $\pm$ 0.22 & -7.14 $\pm$ 2.12 & 0.92 \\
\hline
\end{tabular}
\centering
\caption{The total main sequence fitting parameters, irrespective of morphology, as a function of redshift range.}
\label{tab:main_sequence}
\end{table}

\subsection{Passive and Star Forming from UVJ Diagrams}


As noted in papers previously \citep[e.g.,][]{Williams2009, Ownsworth2016}, the position of galaxies at high redshift in a so-called UVJ diagram can reveal their star forming nature.  Within this space there is a region where passive galaxies are found, while we also find systems that are star forming are found in a lower part the diagram.   Whilst this may seem redundant to the star formation analyses we preform in the last section, we carry out this UVJ analysis as a complement to that work and also a way to better understanding the physical nature of morphology and how it correlates with current stellar population properties. In principle, the UVJ gives some idea of the existing stellar population and whether the galaxy is `passive' or not and using colours we have a longer time-scale in which to determine this than compared to the star formation rate which is more of an instantaneous measurement when the galaxy is observed. 

The colours for the UVJ analyses here are from the catalog provided by \citep[][]{Stefanon2017}.  
 These colours are derived from a BC03 SPS template, and an exponentially declining SFH; with $\tau$ = [ 0.1, 0.3, 0.6, 1, 2, 3, 5, 9, 15 Gyrs]. Both Calzetti and SMC dust extinction laws are implemented, with the code deciding the extinction model which gives the better fit. No nebular emission is added, additionally an array of metallicities is used, including a subset of models with young ages and low metallicities. See \citet{Stefanon2017} and \citet{Mobasher_2015} for greater discussion of how these colours are synthesised.

\begin{figure}
    \centering
    \hspace{-1cm}
    \includegraphics[width=1.15\linewidth]{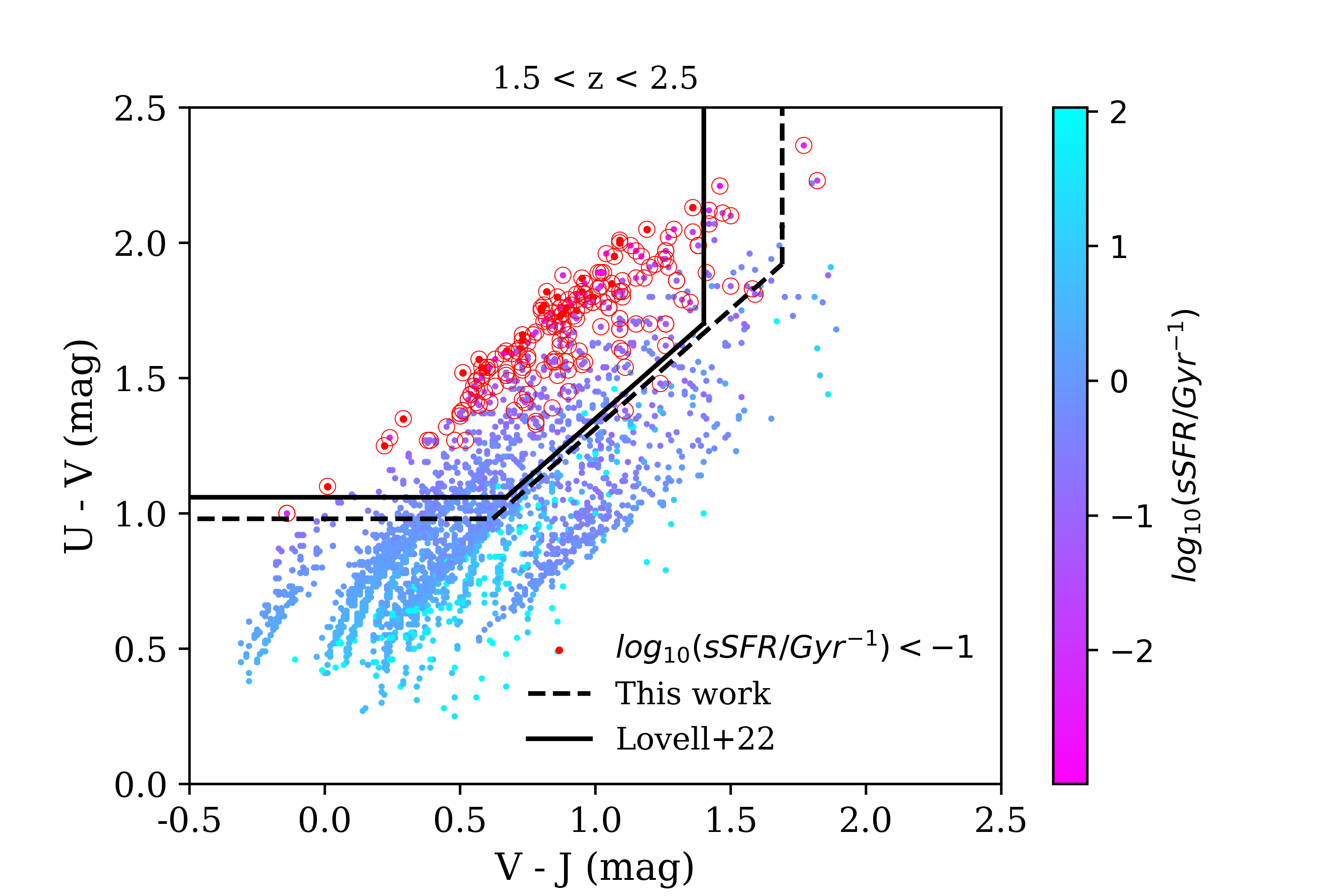}
    \caption{UVJ colours and diagram which are used to define the population cuts in the 1.5 < z < 2.5 bin based on the star formation rate of these galaxies. Galaxies are colour-coded by sSFR (as measured in Gyr$^{-1}$), and those which have log(sSFR) < -1 are circled in red.  This is an example redshift bin, other redshifts show similar results and trends.}
    \label{fig:UVJ_SSFR}
\end{figure}

We use the full CANDELS EGS catalog to define the redshift dependent cuts, identifying the galaxies in which the sSFR fell below a threshold.  We therefore compare the position of galaxies in the UVJ plane with those which are passive based on their low sSFR.
The threshold for finding passive galaxies was taken to be $-1$ Gyr$^{-1}$, as plottedin Figure~\ref{fig:UVJ_SSFR}. The selection region for quiescent galaxies is defined as,

\begin{multline}
    U-V > 1.19 - 0.07(1+z) \\
    V-J < 1.93 - 0.07(1+z) \\
    U-V > 0.88(V-J) - 0.014(1+z) + 0.476 \\
\end{multline}

\noindent These cuts share the same gradient as in the work done by \citet{Lovell2022}, although we introduce a redshift dependent V$-$J bound, accounting for the migration of the galaxies into more positive values of U$-$V and V$-$J towards the present.
Once these cuts are defined for each redshift bin in the range $ 1.5 < z < 7 $, we compare the morphologically classified regions to this.  Overall, however, it is important to note that we find a very good agreement between the sSFR of galaxies and their position in the UVJ colour space.  Galaxies which are defined as passive due to their position in this space have the lowest sSFR values (Figure~\ref{fig:UVJ_SSFR}). In fact, we find that it is the galaxies which are the bluest in both colours, shown in the bottom left part of the diagram, are most likely to have a high sSFR.  Thus, we can use this diagram with some confidence to understanding the past history and the state of the stellar populations within galaxies for different morphological types.

\begin{figure*}
    \centering
    \includegraphics[width=1\linewidth]{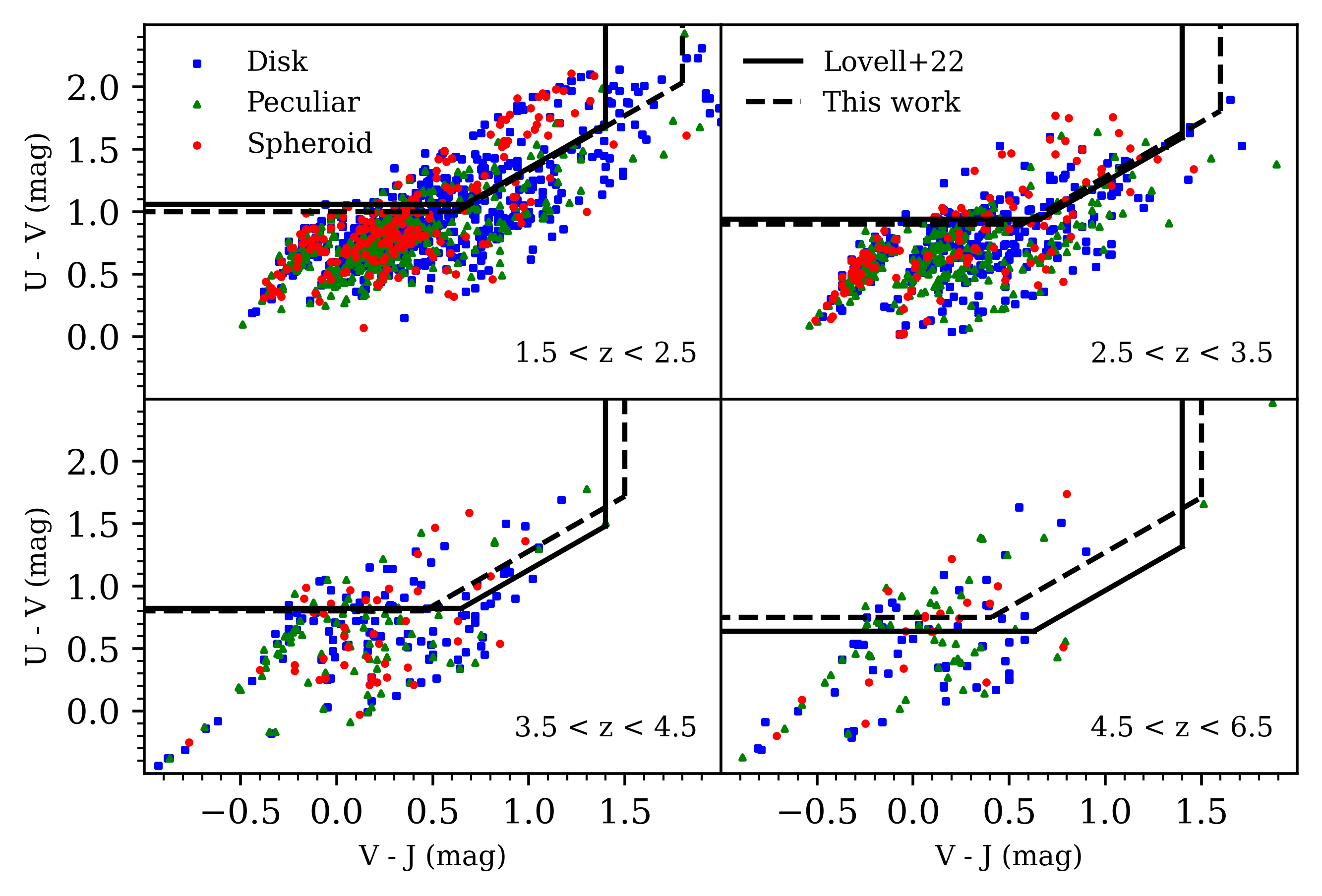}
    \caption{A division of UVJ diagrams for our subset of morphologically classified galaxies up to redshifts $z = 6.5$. Redshift dependent cuts derived in both this work, and that done by \citet{Lovell2022} are shown; dividing the population into passive and star-forming regions. No true bi-modality can be seen at these high redshifts, and the two regions can be seen to be populated by a varying amount at all morphologies. Although a lower proportion of peculiars are seen in the passive domain at late-times, at early-times this proportion can be seen to increase. We see surprisingly very little correlation between morphological type and the location of galaxies on these UVJ diagrams. }
    \label{fig:UVJ_morph}
\end{figure*}

 When we apply the UVJ cut method in our sample as labelled by morphological type, we obtain Figure. \ref{fig:UVJ_morph}, whereby the morphologically classified subset is denoted within this space. By doing this we can determine how  UVJ space is morphologically dependent, finding that there is very little to no correlation between galaxy morphology and the position in the UVJ space.    This is a sign that at the highest redshifts, there is no obvious correlation between a morphological type and the age of its most recent stellar population, consistent with Section 3.4.   This morphological treatment does change however as we go to lower redshifts, whereby the Hubble sequence is established.  

 It is also worth remarking that we find very few of the spheroid types in the passive region of the UVJ space.  In fact at $z > 3.5$ we find that there is no correlation between UVJ colour and the morphological type. 
 While at the lower redshifts, we do see at least some division of types.  In general, we find that the spheroid types are at the blue end of both colours, while the disk galaxies are redder.  We also see a significant overlap of the position of the spheroids with those of the peculiars, suggesting perhaps some association.    This is consistent with the findings of the main-sequence, whereby we find that a significant fraction of the spheroids are located in the starburst or main-sequence of star formation region of this parameterisation.   This is due to the fact that these spheroids are actively forming.  These high-z systems could be the ancestors of ellipticals today, but in a formation mode from some type of merging activity or otherwise a collapse of gas into a small volume. In the next sections we directly investigate the sSFR patterns for these galaxies.

\begin{figure*}
\centering
\includegraphics[width=1.0\columnwidth]{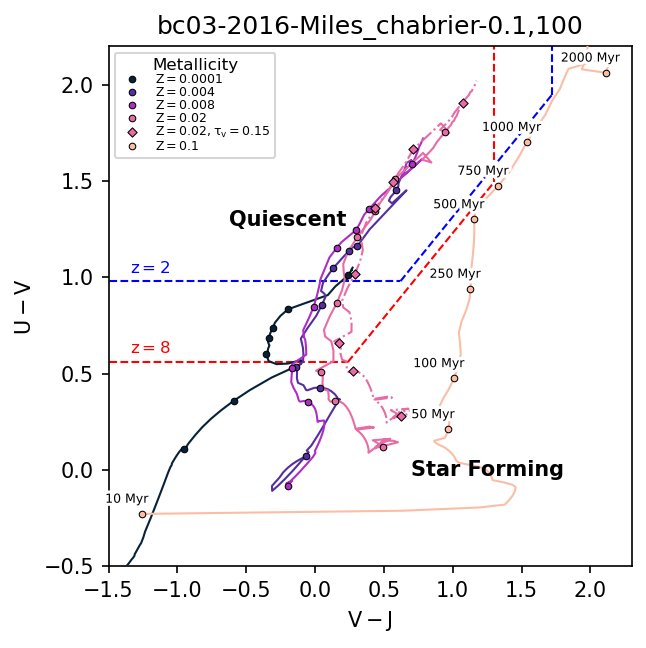}
\includegraphics[width=1.0\columnwidth]{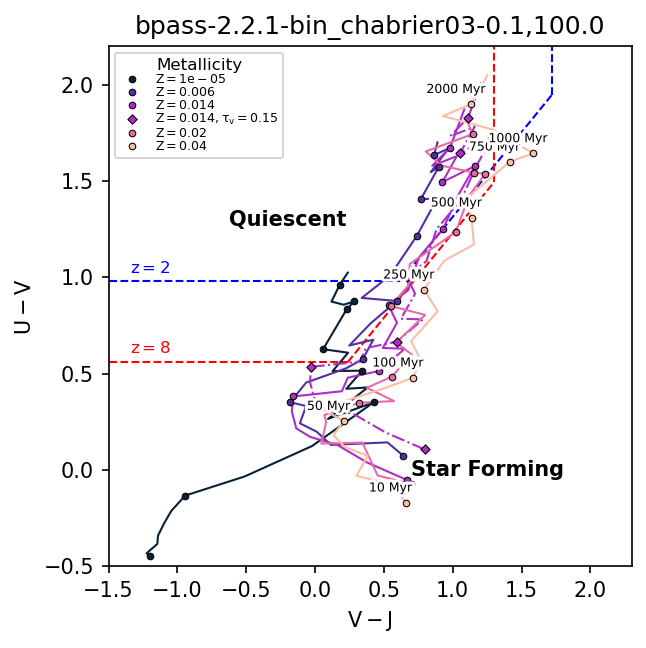}
\caption{The UVJ diagrams, but now showing models of where galaxies of different metallcities (Z) evolve in terms of U-V and V-J colours from 10 Myr to 2Gyr.  On the left we show BC03 \citep[][]{Bruzual2003} with 2016 Miles Stellar population synthesis models, with a Chabrier IMF; on the right panel we show BPASS models \citep[][]{Eldridge2017} also with a Chabrier IMF. Most models show the intrinsic stellar spectra, with no dust included. The markers are (from the bluest, bluest point redward) 10, 50, 100, 250, 500, 750, 1000 and 2000 Myr.   Also shown are the $z =2$ and $z=  8$ quiescent defined regions which we use in Fig~\ref{fig:UVJ_morph} at different redshifts.}
\label{fig:uvj_models}
\end{figure*}

 We can use the models of stellar population evolution shown in Figure~\ref{fig:uvj_models} to determine the likely time-scales in which at least the spheroid systems remain below the passive region on this diagram.  We show in this figure models from \citet{Bruzual2003} with 2016 Miles Stellar population synthesis models, with a Chabrier IMF and BPASS models \citep[][]{Eldridge2017} with the same Chabrier IMF and stellar mass range from 0.1-100 \solm. The markers on the plots are (from the bluest, bluest point redward) 10, 50, 100, 250, 500, 750, 1000 and 2000 Myr. We also include one dusty galaxy model at solar metallicity.  
 
 If we examine the location of the spheroids in this space, we find that they have a stellar population with an age of about 100 Myr, assuming a solar type metallicity.  These same models show that these galaxies would be within the passive region of UVJ space after another few hundred million years.  In this sense the UVJ diagram is revealing that the star formation activity is more prolonged than the morphological evolution for these galaxies, assuming that these are transiting from peculiars into spheriods, a topic we discuss in more detail in the discussion section.

\subsection{Specific star formation rate evolution}

Star formation is the derivative of the stellar mass within galaxies. 
It gives us an idea of the rate at which stars in galaxies are forming at a given instance. 
Another approach we use is investigating the specific star formation rate (sSFR), which gives the relative amount of star formation for an already formed amount of stellar mass.  
For a given observed star formation rate (SFR), the specific star formation rate is simply defined and measured as sSFR = SFR / M$_{*}$.  In our notation this is the derivative of the change in mass divided by the integral of the star formation rate over all of time.   The observed distributions of star formation rates with redshifts are asymmetric, as measured SFRs span many orders of magnitude. 
We report the median sSFR with associated error bars as the 16th and 84th percentile values, as to enclose a 68\% confidence region of the measurement.

\begin{figure}
    \centering
 \hspace{-0.5cm}   \includegraphics[width=1.05\linewidth]{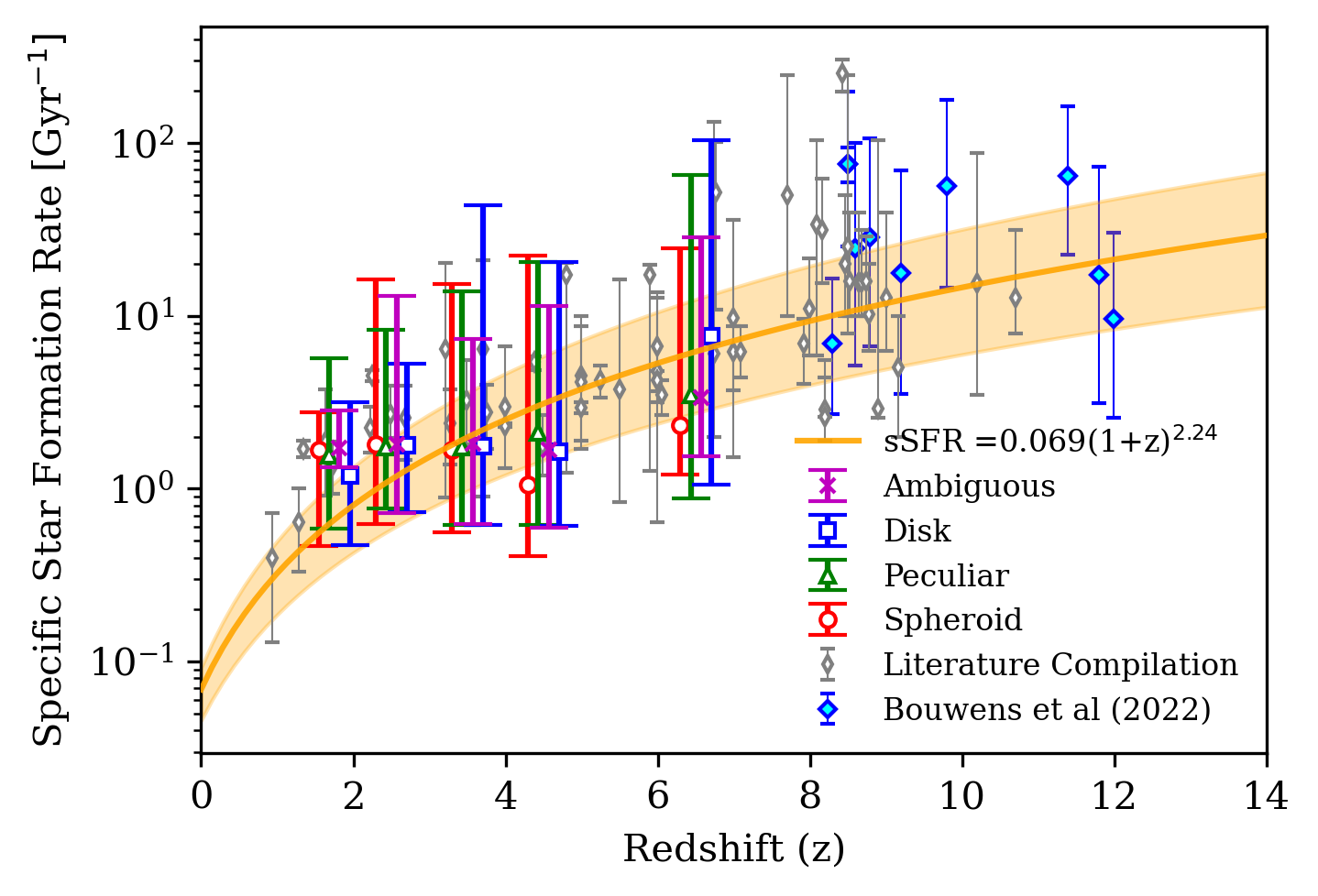}
    \caption[]{The median specific star formation rates for galaxies grouped by morphology as a function of redshift.  The distribution of the sSFR values for the different morphological types are also shown plotted as different colours and symbols.  Also plotted for reference are individual sSFRs from previous work.}
    \label{fig:sSFRtot}
    
\end{figure}

In a sense this is an indication of how important the ongoing star formation rate is within a galaxy compared to the past integrated star formation rate, as measured by the stellar mass.    
We then take the average for the sSFR for each galaxy morphological type at each redshift and plot the results in Figure~\ref{fig:sSFRtot}.

We model the evolution of this sSFR evolution through redshift by a simple power law of the form,

\begin{equation}
    \left(\frac{\textrm{sSFR}}{\textrm{Gyr}^{-1}} \right) = a(1+z)^b
    \label{eqn:sSFR-z}
\end{equation}

\noindent where the parameters $a$ and $b$ are free and fit parameters. We calculate these free parameters for the entire population within the survey and further investigate this fit in relation to the differing morphological and stellar mass bins.

Comparing the individual morphologies to the power law relation found for the entire population, we see no distinction in sSFR in terms of morphological type, as can be seen in Figure~\ref{fig:sSFRtot}.  This implies that the construction of new stellar mass in galaxies during the $1.5 < z < 7$ epoch does not depend on the overall morphology.  This strongly implies that there are several ways in which the star formation in a galaxy can be triggered with different effective morphologies and resulting structures.   It also may imply that galaxy morphology is established first and the star formation occurs within this structure without dramatically changing the overall appearance of a galaxy.   This issue is addressed in more detail in \citet{Ormerod2024}.

\begin{figure}
    \centering
    \includegraphics[width=1.05\linewidth]{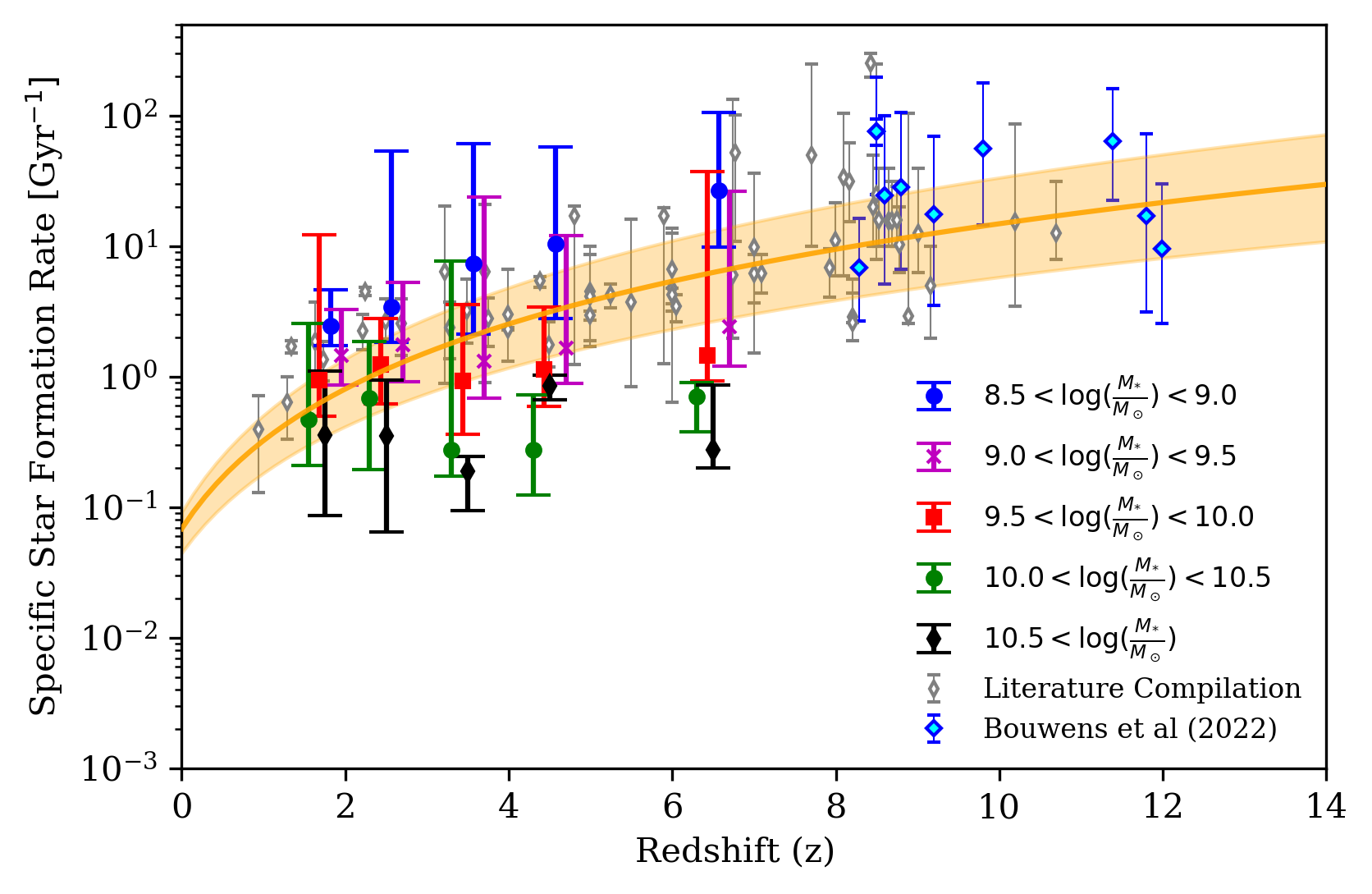}
    \caption{The specific star formation rates for galaxies as a function of stellar mass.  The symbols are the same as in Figure~\ref{fig:sSFRtot}. However, in this instance we have plotted our sample in terms of stellar mass cuts as opposed to morphological type.  A much stronger dependence on stellar mass is seen.}
    \label{fig:sSFRmass}
\end{figure}

Probing the form of this power-law for galaxies of varying stellar masses, however, yields a more diverse array of free parameters, and one that correlates with the value of the stellar mass.
This measurement shows that there is a strong correlation between the evolution of the sSFR and the stellar masses of galaxies (Figure~\ref{fig:sSFRmass}) which is tabulated in Table~\ref{tab:sSFRfits}.  {\rm This evolution reflects a tendency for the higher mass galaxies to drop in sSFR at lower redshifts, while the lower mass galaxies still tend to contain a high star formation rate.  In fact, we find that the fitted parameters for the sSFR evolution finds significant differences between the lowest mass galaxies and the highest in terms of the normalisation parameter $a$, while the slope for all types, the $b$ value, is quite similar. } This is another indication of the so-called galaxy downsizing, whereby the highest stellar mass galaxies decline in their star formation rate and AGN activity before lower mass galaxies \citep[e.g.,][]{Bundy2006a}. We discuss these results in the discussion section along with the other findings of the paper to develop hypotheses for how galaxy formation is driven at early times in the universe's history. 

\begin{table}
    \centering
    \begin{tabular}{c|c|c|c}
    \hline
    \ Stellar Mass & $a$ [Gyr$^{-1}]$ & b & $\chi^2_{\textrm{red}}$\\
    \hline
    $8.5 < \log_{10}\left(\frac{\textrm{M}_*}{\textrm{M}_{\odot}}\right) < 9.0$ & 0.26 $\pm$ 0.18 & 2.41 $\pm$ 0.28 & 3.3 \\
    $9.0 < \log_{10}\left(\frac{\textrm{M}_*}{\textrm{M}_{\odot}}\right) < 9.5$ & 0.11 $\pm$ 0.09 & 2.36 $\pm$ 0.27 & 11.8 \\
    $9.5 < \log_{10}\left(\frac{\textrm{M}_*}{\textrm{M}_{\odot}}\right) < 10.0$ & 0.044 $\pm$ 0.034 & 2.36 $\pm$ 0.27 & 3.7 \\
    $10.0 < \log_{10}\left(\frac{\textrm{M}_*}{\textrm{M}_{\odot}}\right) < 10.5$ & 0.007 $\pm$ 0.004 & 2.34 $\pm$ 0.26 & 15.3 \\
    $10.5 < \log_{10}\left(\frac{\textrm{M}_*}{\textrm{M}_{\odot}}\right)$  & 0.007 $\pm$ 0.003 & 2.39 $\pm$ 0.27 & 47.6 \\
    \hline
    \end{tabular}
    \caption{The power-law fitting parameters for $\log_{10}$[sSFR] values as a function of redshift for various stellar mass ranges of galaxies. As shown, the slopes (the $a$ parameter) of these fits change with redshift, getting progressively steeper at the highest stellar masses. }
    \label{tab:sSFRfits}
\end{table}

\subsection{Galaxy Size and Internal Stellar Mass Density}

Using the galaxy sizes and the masses we have calculated for these objects, we investigate the relation between galaxy morphology/structure and quenching or the ending of star formation. 
There are reasons to think that galaxy density, or rather, galaxy mass divided by the radius (essentially the stellar potential) correlates strongly with the quenching in a galaxy.  
This has been seen at redshifts $z=0-2$ \citep[][]{Bluck2023}, and is related to the idea that these systems are predicted to host higher central supermassive black holes, which facilitates the quenching of galaxies \citep[][]{Piotrowska2022, Bluck2023}.

\begin{figure}
    \centering
    \includegraphics[width=1.1\linewidth]{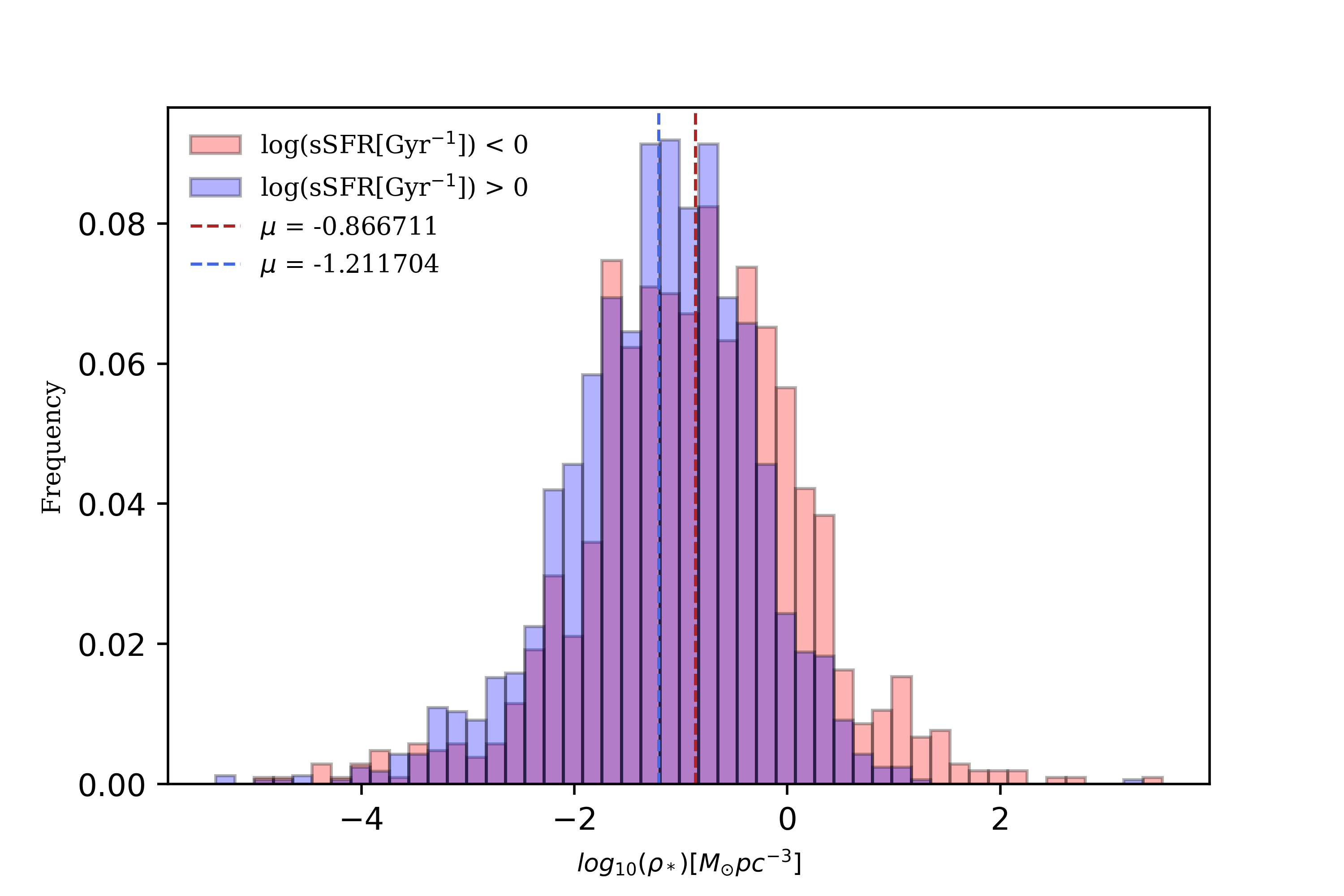}
    \caption{The distribution of internal mass density for our galaxies and how these depend on the sSFR values, shown as histograms.  As can be seen those objects with the highest densities have the the lowest specific star formation rates at the lower values.}
    \label{fig:sSFRhistogram}
\end{figure}

If we examine the distribution of mass density $\rho_*$ values in terms of the sSFR in units of Gyr$^{-1}$ we find that the lower sSFR galaxies tend to have a higher internal mass density, as seen in Figure~\ref{fig:sSFRhistogram}.  This relation is not however perfect, and it is clear that some galaxies with low sSFR rates are in low mass density systems. It appears that the stellar potential may do better at predicting quiescence \citep[e.g.,][]{Bluck2024}.  This is a sign, along with the mass as we have already discussed, which indicates what is likely producing the truncation of star formation in our sample of galaxies. Thus, whilst we find that mass is a factor, we find that as well the density is also a factor in producing a change in the star formation in galaxies.  Although we find that even though overall morphology is not a critical factor in determining the star formation rate and quenching of galaxies, mass density and the stellar mass of a galaxy is proportional to the quenching effect.  We discuss the reasons this might be the case in the discussion section of this paper (\S 4).

\section{Discussion}

There are several ways in which we can measure how galaxy formation and evolution occur.   With the advent of JWST we now have the ability to determine how galaxy morphology is part of this process up to the earliest times where galaxies can be found. This is due to the recent morphological catalogue of \citet[][]{Ferreira2022} whereby a new direction can be taken as we have reliable visually-classified morphologies for galaxies at these early times.  The question we address here is what role morphology plays in the formation of galaxies at $z < 7$ and whether there is a correlation between these morphologies and the way that galaxies have, and are, formed.  We use the star formation rate as a proxy for the ongoing formation rate of galaxies, and the stellar mass as the integrated history of all mass assembly.  While we know that mergers are important at this epoch and may dominate mass assembly \citep[e.g.,][]{duncan2019}, we are not able to measure this directly, although we can get some idea of its presence through morphology.  

The summary of our results is such that we find that galaxy morphology does not appear to be a critical aspect for the formation of galaxies, at least in terms of their star formation rates at $3 < z < 7$.  This means that there are systems which have low and high star formation rates at every visible morphology that we classify our systems into.  It is therefore not the case that only peculiar or disk like morphologies are those that contain a high star formation rate -- we can see that galaxies we classify into spheroid-types also have a high star formation rate quite often in the early universe.   We in fact see very little correlation between properties, except for the trends of sSFR with redshift at different stellar masses, which are significantly different.  Whether these objects are true spheroids, or forming spheroids, is another question.  What is likely shows is that morphology is set early in a galaxy and events such as star formation do not significantly alter these structures.  Mergers may do so and these will be examined in future studies.   

From previous results, mostly from HST, we know that there is an apparent correlation between the redshifts of galaxies and their structures as given by the redshift-morphology relation  \citep[e.g.,][]{Conselice2014a, Mortlock2013}.  In addition, processes such as mergers can be seen up to these high redshifts through using pairs of galaxies \citep[e.g.,][]{duncan2019}, and these mergers can have an influence on the galaxy structures present, as well as their measured star formation rates.  Furthermore, now that we have this new parameter space of morphology for distant galaxies we can test whether and how correlations exist that might suggest causality, or to lead to further investigations of why and how galaxy structure and formation of mass in galaxies might be correlated \citep[][]{Conselice2006b, Kelvin2014, Bluck2023}. 

In this paper, using JWST observations we find that galaxy morphology, at least as classified by eye into the classic spheroid, disk, and peculiar types is not a dominating feature in the formation of galaxies at $z > 3$.  Whilst we do find a strong differential in morphological types at lower redshifts at $z < 1$ this difference does not appear to exist at higher redshifts as strongly.    In fact, we find that galaxy mass functions are clearly different for galaxies that are classified as disks vs. those which are classified as ellipticals at lower redshifts.    We find that galaxy morphology does not correlate strongly with specific star formation rate, or with the fraction of passive galaxies as seen at high redshift.  Essentially, we are finding that galaxy structure and galaxy formation processes are largely decoupled to some extent at high redshifts, as has also been previously claimed for deep HST observations with WFC3 \citep[][]{conselice2011}. We can measure the time-scales for this, given the SFR measurements and the colours of our various morphological types.  First, we know that the SFR we calculate through UV emission has a time-scale of roughly 100 Myr, thus if some galaxies are formed in bursts and little subsequent formation, such as ellipticals and spheroids, then we would expect them to have low SFR and sSFRs after 100 Myr.  Likewise, we can investigate the colours of our galaxies within the UVJ diagram and use stellar population models to put limits on the morphological and star formation time-scales.

We show the result of this in Figure~\ref{fig:uvj_models}, which we have previously discussed when introducing the UVJ results. As shown there, the time-scale for the spheroids to move from the star forming region into the passive one is several hundred Myr.  This would imply that by the time these morphological types are at $z = 1.5$ we would find them in the passive region, thus establishing the traditional Hubble sequence of finding a correlation between galaxy morphology (appearance) and star forming properties.  These galaxies must therefore form their morphology and structure more rapidly than ending their star formation. If this star formation is truncated by feedback, then this feedback occurs in a longer period of time than the processing resulting in the morphological establishment of these systems \citep[][]{Hopkins2010}.  We also find that the peculairs are in the similar region of the UVJ space as shown in Figure~\ref{fig:UVJ_morph}.  This may indicate that these galaxies are rapidly evolving into spheroids whilst still undergoing star formation. As these effects are decoupled and the dynamical effects of changing the morphology can occur within $< 0.5$ Gyr \citep[][]{Whitney2021}  it seems likely that this is why we are finding spheroids in the part of the UVJ space at $z > 1.5$.   The high merger rate for galaxies suggests that this is the reason for the delayed formation of passive spheroids \cite[e.g.,][]{duncan2019, Whitney2021}. 

\begin{figure*}
    \vspace{-2cm}
    \centering
    \includegraphics[width=\linewidth]{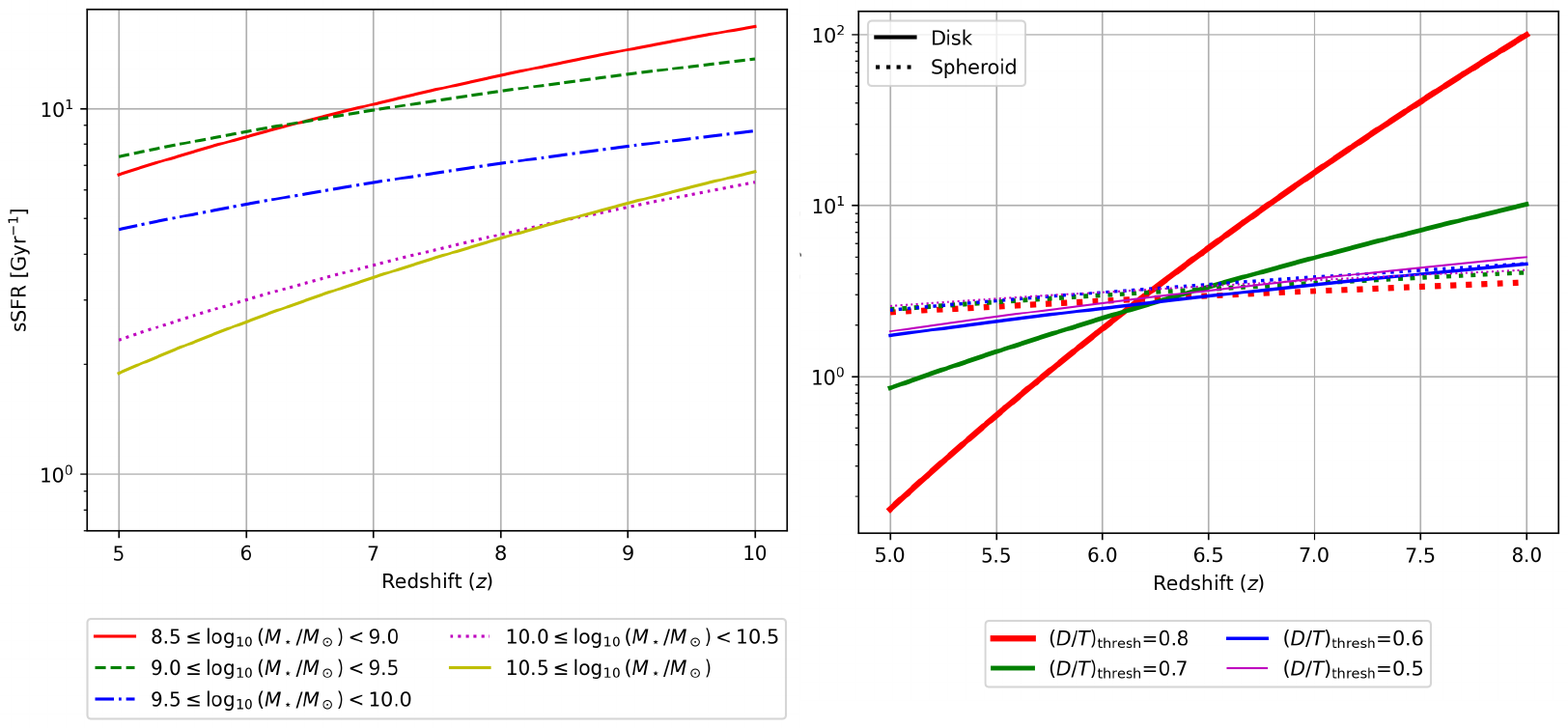}
    \vspace{-2.5cm}
    \caption{The results of theory for measuring these quantities as taken from the FLARES simulation from \citep[][]{Roper2022} . As can be seen we find a stronger relation between the stellar mass selection for the sSFR evolution, than we do in the morphological cuts, here taken as bulge to total ratios, which correlate to some degree with morphological types \citep[][]{Ormerod2024}.}
    \label{fig:theory}
    \end{figure*}

Throughout the redshift bins we have studied in this paper at $z < 7$, we also investigate the distribution of SFR within each morphological type, finding similar conclusions.  Stellar mass however strongly correlates with the sSFR, which we can see as well when comparing our observations to theory from the FLARES simulation \citep[e.g.,][]{Lovell2021,Roper2022}, as shown in Figure~\ref{fig:theory}, where galaxies with stellar masses > $10^{10} \textrm{M}_\odot$ were decomposed into a bulge and a disc component based on the method introduce by \cite{Irodotou2021}. 

\begin{figure}
    \centering
\hspace{-0.5cm}    \includegraphics[width = 1.1\linewidth]{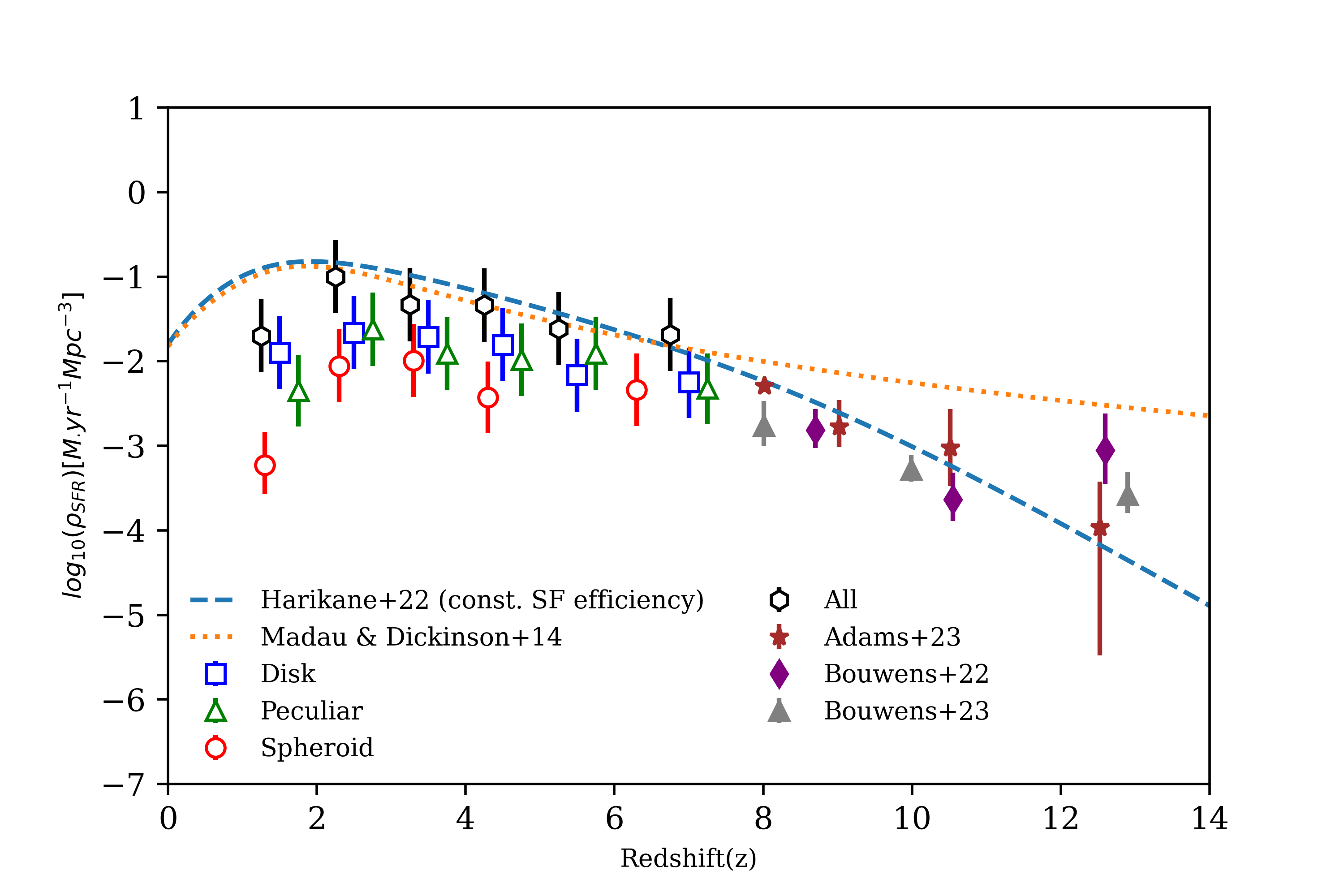}
    \caption{The SFR density of the universe at different epochs of time. We obtain this for each morphology by integrating the corresponding SFR functions, that is number of galaxies at a given star formation rate, in redshift bins. For comparison we include data points from \citet{Adams2023,Bouwens2023,Bouwens2022}. The blue (dashed) line is the constant star formation efficiency model from \citet{Harikane2023}, whereas the orange (dotted) line is the double power law derived by \citet{Madau2014}.}
    \label{fig:sfr_density_morph}
\end{figure}

Another way to investigate these questions is to probe how much star formation is within each morphological type as a function of redshift.  This was presented to some degree already in the study of overall morphology in \cite{Ferreira2022b}.  Here we look at this in a very broad way.  As shown in Figure~\ref{fig:sfr_density_morph}, there is a difference in the relative contribution of galaxy types to the star formation rate density.  As can be seen, there is no particular morphological type that dominates the star formation of the universe at $z < 7$.  Again, this demonstrates that there are different modes of producing star formation, at least assuming these are reflected in the morphologies of galaxies, but there is a range of these structures at the high end of the star formation rate distribution at $z > 3$.  If anything we see that there is a divergence of the star formation rate for spheroids at $z < 2$, which may signify that this type of star formation, producing a spheroidal structure, is petering out at lower redshifts.  One obvious explanation is that these spheroidal morphologies are formed from mergers of galaxies, and thus when the merger rate declines \citep[][]{duncan2019, Conselice2022} steeply at lower redshifts, this results in fewer compact elliptical like galaxies forming.    

Likewise, we see that the galaxies classified as disks dominate the fraction of ongoing star formation rate at all redshifts, with this especially being the case at the lower redshifts (see also \citep[][]{Ferreira2022b} where this is also discussed in detail). This means that physics of galaxy and star formation is likely dominated by whatever processes triggers this in disk galaxies.  The answer to this is beyond the scope of this current paper.  This is likely either from gas accretion or perhaps merging galaxies that produces a disk galaxy as a final product, although this must happen very early in the universe.   Some analyses show that within a mass selected sample it is indeed gas accretion which dominates the star formation \citep[e.g.,][]{Conselice2012a}, and it is likely this mode which keeps these galaxies from quenching.     However, more studies are needed to investigate the ways in which star formation is triggered and sustained in disk galaxies at high redshifts to indeed investigate this idea.

However, what we are also interested in is inferring how star formation is quenched in distant galaxies. This is a separate question to that of how star formation is triggered.  What we find is that stellar mass is a factor in this, but we also find that galaxy density, as defined as the stellar mass divided by the radius or size squared tends to be a factor in which galaxies have a lower star formation rate.  This implies that galaxies which are denser in matter have a lower star formation rate.  This is also seen for nearby galaxies \citep[e.g.,][]{Bluck2024}, and is likely, or probably, a result of having more super massive black holes in these lower star forming systems.  This is an example of quenching formed by 'mass' or by feedback from AGN, something that we investigate more in \cite{Bluck2024}

Another issue that our results imply is that galaxy morphology in the traditional Hubble division of spheroid/disk/peculiar is not sufficient to trace the major formation mechanisms occurring at high redshifts. It is possible, or perhaps even likely, that there are other classes of galaxies that can be developed at higher redshifts that correlate better with ongoing physical processes driving galaxy assembly and the quenching of galaxies.  \cite{tohill2023} in fact suggests using JWST imaging that there are other groups in which galaxies should be classified based on unsupervised machine learning.  Time will tell how well this new approach is towards understanding the evolution and establishment of galaxy structure and morphology in the universe. 

\section{Conclusions}

As part of our series of papers using EPOCHS data, we describe the dependence of morphology on the star formation and stellar mass assembly history of galaxies. 
In this paper we investigate the morphological evolution of galaxies of various types and compare this with their stellar mass, stellar populations, and star formation rates.  Our primary findings include:

I. The stellar mass function for these galaxies evolves such that the disk galaxies have a very similar pattern of stellar mass distributions at all redshifts $z < 7$, whereas the peculiar galaxies decrease in average stellar mass over this time.  The spheroid galaxies have the lowest number densities and flattest $\alpha$ slopes at the lower redshifts.

II. The specific star formation rate (sSFR) for galaxies as a function of morphology is very similar between the different morphological types.  That is, the relative amount of star formation in a galaxy at $z < 7$ does not strongly depend upon its morphology --  we find that spheroids, disks and peculiars all appear to have the same amount of average star formation, for a given stellar mass. This signifies that star formation can be produced in different ways in early galaxies and these can produce morphologies that at least appear to resemble later galaxy morphological types, or are established in those forms whilst undergoing star formation events that do not alter the overall morphologies. 

III.  We find very little correlation between the morphology of a galaxy and its star formation rate, both past, and current.  We find that all morphological classes of galaxies can have high star formation rates, with no significant difference in the fractions of galaxies of various types found to be in different morphological states. Most notably we find that galaxies classified by eye as spheroids have a range of star formation rates, and thus they are not perfect analogs to the ellipticals that we find in the nearby universe, which are for the most part passive galaxies.

IV. We find that stellar mass is the dominant property which determines when a galaxy will start to lower its star formation rate by comparing the sSFR of galaxies as a function of mass and redshift. This is further seen in simulations of galaxy formation. We further show that the mass density within galaxies has an important effect on the star formation rate, with the highest density objects tending to have the lowest star formation rates, although there is not a 1:1 correlation here.  

Overall our results suggest, within the milieu of stellar mass measurements being correct, that there is a strong trend during all of cosmic time for the mass of galaxies to strongly depend upon their formation histories. These formation histories of galaxies are not reflected in any obvious way with the underlying morphologies of galaxies, such that we cannot use morphology to predict the star formation rate, which is possible to some degree within the nearby universe \citep[e.g.,][]{Conselice2006, Kelvin2014}.  Morphology therefore might be a very stable feature of galaxies, that are imprinted very early in their history and outside of major dynamical events or accretion, they retrain the same morphology over most of cosmic time. 

In the future, efforts to quantify the amount of light in these galaxies and how they are distributed will help resolve some of these questions.  This sample is also investigated in companion papers include \citet{Ormerod2024}.  By examining features such as the Sersic index as well as the radius of these galaxies, and also their non-parametric morphologies, we obtain a better idea of how galaxy structure influences the star formation and mass assembly of distant galaxies.  This is also the case for methods such as those that utlise machine learning \citep[][]{tohill2023}, finding new classes of galaxies that can then be studied for their properties as a function of redshift as well as how these evolve with time.

\vspace{-0.5cm}
\section*{Acknowledgements}

 We acknowledge support from the ERC Advanced Investigator Grant EPOCHS (788113), as well as a studentship from STFC. LF acknowledges financial support from Coordenação de Aperfeiçoamento de Pessoal de Nível Superior - Brazil (CAPES) in the form of a PhD studentship. DI acknowledges support by the European Research Council via ERC Consolidator Grant KETJU (no. 818930). This work is based on observations made with the NASA/ESA \textit{Hubble Space Telescope} (HST) and NASA/ESA/CSA \textit{James Webb Space Telescope} (JWST) obtained from the \texttt{Mikulski Archive for Space Telescopes} (\texttt{MAST}) at the \textit{Space Telescope Science Institute} (STScI), which is operated by the Association of Universities for Research in Astronomy, Inc., under NASA contract NAS 5-03127 for JWST, and NAS 5–26555 for HST.

This research made use of the following Python libraries: \textsc{Astropy} \citep{astropy2022}; \textsc{Morfometryka} \citep{ferrari2015}; \textsc{Pandas} \citep{pandas}; \textsc{Matplotlib} \citep{Hunter:2007}; \textsc{photutils} \citep{larry_bradley_2020}

\section*{Data Availability}
The specific observations analyzed can be accessed via \url{https://doi.org/10.17909/xm8m-tt59}, and the visual classifications from \citep{Ferreira2022b} are publicly available at \url{https://github.com/astroferreira/CEERS_EPOCHS_MORPHO/}.



\bibliographystyle{mnras}
\bibliography{paper} 






\bsp	
\label{lastpage}
\end{document}